\documentclass[aps,prd,onecolumn,groupedaddress,,showkeys,nofootinbib,amssymb]{revtex4}
\usepackage[T1]{fontenc}
\usepackage[latin1]{inputenc}
\usepackage{graphicx}
\usepackage[english]{babel}
\usepackage{graphicx}
\usepackage{bm}
\usepackage{amsmath}
\usepackage{amssymb}
\usepackage{amsfonts}
\usepackage{epsfig}
\usepackage{colordvi}
\usepackage{color}
\usepackage{appendix}
\newcommand{\lp}{\left(}
\newcommand{\rp}{\right)}

\begin{document}
%\title{Newtonian limit for $f(R, \mathcal G)$-modified gravities}
\title{Newtonian, Post Newtonian and Parameterized Post Newtonian limits of  $f(R, \mathcal G)$ gravity}

\author{Mariafelicia De Laurentis$^{1,2}$, Antonio Jesus Lopez-Revelles$^{3}$}

\affiliation{$^{1,2}$\it Dipartimento di Fisica, Università
di Napoli {}``Federico II'', Compl. Univ. di
Monte S. Angelo, Edificio G, Via Cinthia, I-80126, Napoli, Italy,\\
 INFN Sez. di Napoli, Compl. Univ. di
Monte S. Angelo, Edificio G, Via Cinthia, I-80126, Napoli, Italy}

\affiliation{$^{3}$ \it  Consejo Superior de Investigaciones
Científicas, ICE/CSIC-IEEC, Campus UAB, Facultat de Ciencies, Torre
C5-Parell-2a pl, E-08193 Bellaterra (Barcelona) Spain}

\date{\today}
\begin{abstract}
We discuss in detail the weak field limit of $f(R,\mathcal{G})$ gravity taking into account analytic functions of the Ricci scalar $R$ and the Gauss-Bonnet invariant  
$\mathcal{G}$. Specifically,  we 
develop, in metric formalism,  the Newtonian, Post Newtonian and
Parameterized Post Newtonian limits  starting from   general  $f(R, {\cal
G})$ Lagrangian. The special cases of $f(R)$ and $f(\mathcal{G})$
gravities are considered. In the case of the
Newtonian limit of $f(R, {\cal G})$ gravity, a general solution in
terms of  Green's functions is achieved.
\end{abstract}

\maketitle

%%%%%%%%%%%%%%%%%%%%%%%
%%%%%%%%%%%%%%%%%%%%%%%
 \section{Introduction}
 \label{uno}
%%%%%%%%%%%%%%%%%%%%%%%
%%%%%%%%%%%%%%%%%%%%%%%

Any  good theory of physics should  satisfy three main 
 viability criteria that are  self-consistency, completeness, and agreement
with experiments. These have to hold also for theories of gravity like General Relativity (GR). However, despite its successes and  elegance, such a theory  exhibits a number of inconsistencies and weaknesses
that have led many scientists to ask, whether, it is the final theory that
can explain definitively the gravitational interaction. GR disagrees
with a growing number of data observed at infrared scales like cosmological scales.
Furthermore,  GR is not renormalizable, because it presents 
singularities at ultraviolet scales (e.g.  Big Bang singularity, black holes), and it
fails to be quantized as standard quantum field theories. 

 Moreover,  large
amounts of dark matter and dark energy are required to address dynamics
 and structure of galaxies, clusters of galaxies and global accelerated expansion of the Hubble cosmic flow. 
 If one wants to keep  GR and its low energy limit, we must
necessarily introduce these still unknown ingredients that, up to now, seem highly elusive. 
In other words,  GR, from ultraviolet to infrared scales, cannot be the ultimate theory of gravity  even if it addresses 
 a wide range of phenomena. 
 
In order to overcome the above mentioned problems and simultaneously  obtain a semi-classical
approach where GR and its results can be recovered,  Extended
Theories of Gravity (ETGs)  have been recently  introduced
\cite{PhysRepnostro,OdintsovPR,Mauro,Nojiri:2006ri,DeFelice,faraoni}.
These theories are based on corrections and extensions of Einstein's
theory. The effective  Hilbert-Eintein action  is modified by adding higher order
terms in curvature invariants as $R^2$, $R_{\mu\nu} R^{\mu\nu}$,
$R^{\mu\nu\delta\rho}R_{\mu\nu\delta\rho}$, $R\,\Box\,R$ ,
$R\,\Box^{n} R$ or non-minimally coupled terms between scalar fields
and geometry as $\phi^2 R$. The  simplest generalization   is the $f (R)$ gravity where the Hilbert-Einstein action, linear in the Ricci scalar $R$,  is replaced by a generic function  of it
 \cite{buch,ker,bar,mag,Stel}. Such modifications
of GR generate inflationary behaviors which solve a lot of
shortcomings of the Standard Cosmological Model as shown by
Starobinsky  \cite{Star80}, or can explain the flat rotation
curves of galaxies or dynamics of  galaxy  clusters  \cite{annalen}.

Another interesting curvature quantity that should be considered is the Gauss-Bonnet (GB) curvature
invariant   ${\cal G}$ defined below.  This term can avoid  ghost contributions and contribute to the regularization of  the gravitational action   \cite{ghost1, ghost1d, ghost1dd, ghost1ddd, ghost2, ghost2d, ghost2dd, Cal05, DeFelice06, DeFelice06d}).
 Furthermore, in the case in which the Lagrangian density $f$ is a
function of ${\cal G}$,{\it  i.e.}, $f({\cal G})$ it is possible to
construct viable cosmological models that are consistent with local   constraints of GR \cite{fGO, DeTsuji1, DeTsuji2,
Cognola, DeHind, LiMota, Zhou, Uddin}. 

In general,  we can consider
 the most general Lovelock modification of gravity implying curvature and topological invariant, that is 
${\cal L}=f(R,{\cal G})$ \cite{Cognola:2006eg}.
 Beyond the motivations for
considering ETGs, it is
important that we understand their weak-field limit from theoretical and experimental points of view. The Parameterized Post Newtonian (PPN) formalism is the first context for considering  weak-field effects to test their viability with respect to GR. Eddington,
 Robertson and  Schiff  were the first that  rigorously formulated the PPN  formalism and used it in interpreting Solar Systems experiments \cite{rob62,Edd,sch67}.  After   Nordvedt and Will fixed systematically the approach \cite{Nor,tegp}.  
 
 Our aim, here, is to develop the weak field  limit of $f(R,{\cal G})$ considering  the PPN formalism.   Previous results have been developed for $f(R)$ gravity in
  \cite{cap,olmo} where  the Post Newtonian (PN) limit is achieved by using the equivalence with scalar-tensor theories \cite{frst}.
On the other hand, the  PN limit of Gauss-Bonnet gravity has been
developed in \cite{sot} but not for pure  $f({\cal G})$ gravity. Here, we develop
the PPN limit for $ f (R, {\cal G}) $, starting from the field
equations. In doing so,  we restrict our attention to those theories
that admit a Minkowski background. Furthermore, as a special case,
we develop the PPN  limit even for the $f({\cal G})$ and then make a
comparison with the results obtained for $f (R)$. 

The paper is
organized as follows. In Sect.~II, the field equations for $f(R,
\mathcal{G})$-gravity theories are reviewed and the Newtonian,
PN and PPN limits are obtained. In Sect.~III, the Newtonian limit for
$f(R,\mathcal{G})$ gravities achieved in terms of Green's
functions. In Sect.~IV, the weak field limit of the special cases of
$f(R)$ and $f(\mathcal{G})$ modified gravities are, respectively,
studied. Finally, in Sect.~V, we summarize the obtained results  and give some outlooks for the approach.

%%%%%%%%%%%%%%%%%%%%%%%%%%%%%%%%%%%%%%%%%%%%%%%%%%%%%%%%%%%%%%%%%%%%%%%%%%%%%%%%%%%%%%%%%%
%%%%%%%%%%%%%%%%%%%%%%%%%%%%%%%%%%%%%%%%%%%%%%%%%%%%%%%%%%%%%%%%%%%%%%%%%%%%%%%%%%%%%%%%%%
\section{$f(R, \mathcal G)$ modified gravity: the field equations and the Newtonian limit}
%%%%%%%%%%%%%%%%%%%%%%%%%%%%%%%%%%%%%%%%%%%%%%%%%%%%%%%%%%%%%%%%%%%%%%%%%%%%%%%%%%%%%%%%%%
%%%%%%%%%%%%%%%%%%%%%%%%%%%%%%%%%%%%%%%%%%%%%%%%%%%%%%%%%%%%%%%%%%%%%%%%%%%%%%%%%%%%%%%%%%

        This section is devoted to the study of the field equations for $f(R, \mathcal G)$-gravity and their
        Newtonian, PN and PPN limits (for other examples of weak field limit in
        modified gravity theories, see \cite{Capozziello:2010gu,Capozziello:2007ms,Capozziello:2009vr,Stabile:2010zk}).
        The first remarkable characteristic of
        $f(R, \mathcal G)$ modified gravity is that, in this case, we obtain fourth--order field equations, instead of the
        standard second--order ones obtained in the case of GR. This fact is due to the existence of
        some boundary terms that disappear in GR thanks to the divergence theorem, but they remain in other theories,
        as in the case of $f(R, \mathcal G)$ gravity.

        \subsection{Field equations for $f(R,\cal G)$ gravity }
        The starting action for $f(R, \mathcal G)$ gravity is given by:
                \begin{equation}\label{8g1}
                    S=\int d^4x\sqrt{-g} \left\{ \frac{1}{2\kappa^2} f(R, \mathcal{G}) + \mathcal{L}_\mathrm{matter} \right\},
                \end{equation}
            where  $\mathcal{L}_\mathrm{matter}$ is the matter Lagrangian, $g$ is the determinant of the metric, $\kappa^2 = 8 \pi G_N$ is standard gravitational coupling, and $\mathcal G$ is the Gauss-Bonnet invariant,
            defined as
                \begin{equation}\label{8g0}
                    \mathcal G = R^2 - 4 R_{\alpha \beta} R^{\alpha \beta} + R_{\alpha \beta \rho \sigma} R^{\alpha \beta
                    \rho \sigma}.
                \end{equation}
The variation of Eq.(\ref{8g1}) with respect to the metric tensor $g_{\mu \nu}$ gives   the field equations
            for $f(R,\mathcal G)$-gravity that are:
                \begin{equation}\label{8g2}
                    - \frac{1}{2} g_{\mu \nu} f(R, \mathcal G) + f_R (R, \mathcal G) R_{\mu \nu} + g_{\mu \nu} \nabla^2
                    \lp f_R (R, \mathcal G) \rp - \nabla_\mu \nabla_\nu \lp f_R (R, \mathcal G) \rp +$$
                    $$+ 2 f_\mathcal{G}(R, \mathcal G) R R_{\mu \nu} - 4 f_{\mathcal G}(R, \mathcal G) R_{\mu \rho}
                    R_{\nu}^{\ \, \rho} + 2 f_{\mathcal G}(R, \mathcal G) R_{\alpha \beta \rho \mu} R^{\alpha \beta
                    \rho}_{\ \ \ \ \nu} - 4 f_{\mathcal G}(R, \mathcal G) R_{\mu \rho \nu \sigma} R^{\rho \sigma} +$$
                    $$+ 2 g_{\mu \nu} R \nabla^2 f_{\mathcal G}(R, \mathcal G) - 4 g_{\mu \nu} R_{\rho \sigma} \nabla^\rho
                    \nabla^\sigma f_{\mathcal G}(R, \mathcal G) - 2 R \nabla_\mu \nabla_\nu f_{\mathcal G}(R, \mathcal G)
                    - 4 R_{\mu \nu} \nabla^2 f_{\mathcal G}(R, \mathcal G) +$$
                    $$+ 4 R_{\nu \rho} \nabla^\rho \nabla_\mu f_{\mathcal G}(R, \mathcal G) + 4 R_{\mu \rho} \nabla^\rho
                    \nabla_\nu f_{\mathcal G}(R, \mathcal G) + 4 R_{\mu \rho \nu \sigma} \nabla^\rho \nabla^\sigma
                    f_{\mathcal G}(R, \mathcal G) = 2 \kappa^2 T_{\mu \nu}\,.
                \end{equation}
            The trace equation is 
                \begin{equation}\label{8g3}
                    - 2 f(R, \mathcal G) + f_R (R, \mathcal G) R + 3 \nabla^2 f_R (R, \mathcal G) + 2 f_{\mathcal G}
                    (R, \mathcal G) \mathcal G + 2 R \nabla^2 f_{\mathcal G}(R, \mathcal G) - 4 R_{\rho \sigma}
                    \nabla^\rho \nabla^\sigma f_{\mathcal G}(R, \mathcal G) = 2 \kappa^2 T.
                \end{equation}
            In Eq.(\ref{8g2}) and Eq.(\ref{8g3}), the following notation has been used: $\displaystyle{f_R
            (R, \mathcal G) = \frac{df(R, \mathcal G)}{dR}}$ and $f_{\mathcal G} (R, \mathcal G) = \frac{df(R, \mathcal
            G)}{d\mathcal G}$, while $\nabla$ is the  covariant derivative.
The fact that the scalar curvature, $R$, and the Gauss-Bonnet invariant, $\mathcal G$, involves second
            derivatives of the metric tensor $g_{\mu \nu}$ makes of Eq.(\ref{8g2}) and Eq.(\ref{8g3}) fourth-order
            differential equations in  the metric $g_{\mu \nu}$.

        \subsection{ The Newtonian, Post Newtonian and Parameterized Post Newtonian limits}

%           {\color{magenta}\bf The solar system  has weak gravity, and it is the best laboratory where you can make experiments to test several theories of gravity.  Weak- field field, slow motion expansion means to have:
%            \begin{itemize}
%             \item empty and flat space-time at zero order
%           \item the Newtonian corrections of solar system at first order
%           \item post Newtonian corrections to the Newtonian expansion at second order
%
%           \end{itemize}  }

            All the quantities involved in the Newtonian, PN and PPN limits for 
            $f(R, \mathcal G)$ gravity, given can be expanded in powers of $\bar v^2$.
            As  it is well known in the Solar System, the effects of the gravitational field are weak and are well represented by Newton's theory of gravity \cite{will}, 
therefore,  the light rays travel on straight lines at a constant speed and the test particles move with an acceleration
\begin{equation*}
{\bf a}=\nabla U\,,            
             \end{equation*}            
where $U$ is the Newtonian potential produced by a rest mass density $\rho$ according to following relations
\begin{equation*}
\nabla^2 U=-4\pi\rho\,, \qquad U({\bf x},t)=G_N\int d^3 x'\frac{\rho(x',t)}{|x-x'|}\,.
   \end{equation*}
 The hydrodynamic equations for a perfect non-viscous fluid are the usual Eulerian equations     
 \begin{eqnarray*}
 \frac{\partial}{\partial t}+\nabla \cdot (\rho {\bf v})&=&0\,,\nonumber\\
 \rho \frac{d{\bf v}}{dt}&=&\rho \nabla U-\nabla p\,,\nonumber\\
\frac{d}{dt}&=&\frac{\partial}{\partial t}+{\bf v}\cdot \nabla\,,
   \end{eqnarray*}        
   where ${\bf}$ is the velocity of an element of the fluid, $\rho$ is the density and $p$ is the pressure on the element. 
             This means 
             \begin{equation*}
            U\sim \frac{p}{\rho} \sim {\bar v}^2 \sim {\cal O}(2)\,.
            \end{equation*}
%where $U$ is the Newtonian potential, $\bar v$ is the velocity a fluid element.  
Also the derivatives with respect to the time  relative to spatial derivatives are:
\begin{equation*}
\frac{\vert \frac{\partial}{\partial t}\vert}{\vert \frac{\partial}{\partial x}
  \vert} \sim {\cal O}(1).
\end{equation*}
Here we have chosen to set $c=1$.
However, the Newtonian limit is no longer sufficient when we require that, in experiments, accuracies go beyond $10^5$ ({\it e.g.} it is not possible to explain, in the case of Mercury, the additional perihelion advance greater than $5\times10^{-7}$ radians per orbit). 
Thus we need a more accurate approximation that goes beyond the Newtonian approximation, and then  the PN,  PPN limits are considered.  In order to build the PPN limit we need  of expanding in this order of smallness. To recover the Newtonian limit we must develop $g_{00}$ up to ${\cal O}(2)$, while the PN limit requires

\begin{align*}
&g_{00} \qquad \text{up to} \qquad {\cal O}(4)\,,\\
&g_{0i} \qquad \text{up to} \qquad {\cal O}(3)\,,\\
&g_{ij} \qquad \text{up to} \qquad {\cal O}(2)\,,
\end{align*}
where the Latin indices denote  the spatial indices.
 These terms must contains factors as velocity or time derivatives. Furthermore these terms could represent  energy dissipation or absorption. Beyond $ {\cal O}(4)$, modified gravity effects  can take place and give different predictions.
            We expect that it should be possible to find a coordinate system where the metric tensor is nearly equal
            to the Minkowski tensor $\eta_{\mu \nu} = \mbox{diag}(1, -1, -1, -1)$, the corrections being expandable in
            powers of $\bar v^2$. We will consider the following ansatz for the metric tensor:
                \begin{equation}\label{8nl1}
                    \left\{
                    \begin{array}{lll}
                        g_{00} = g_{00}^{(0)} + g_{00}^{(2)} + g_{00}^{(4)} + {\cal O}(6) \ \
                        & \mbox{with} & \ \left\{
                            \begin{array}{l}
                                g_{00}^{(0)} = 1 \\
                                g_{00}^{(2)} = - 2 U
                            \end{array}
                        \right. \\
                        \\
                        g_{0i} = g_{0i}^{(3)} + {\cal O}(5) \\
                        \\
                        g_{ij} = g_{ij}^{(0)} + g_{ij}^{(2)} + g_{ij}^{(4)} + {\cal O}(6) \ \
                        & \mbox{with} & \ \left\{
                            \begin{array}{l}
                                g_{ij}^{(0)} = - \delta_{ij} \\
                                g_{ij}^{(2)} = - \delta_{ij} 2 V
                            \end{array}
                        \right.
                    \end{array}
                    \right.
                \end{equation}
            where $\delta_{ij}$ is the Kronecker delta and $U,V$ are potentials, in particular $V$, following \cite{will}, is define as
            \begin{eqnarray}
           V_i&=&G_N\int d^3 x'\frac{\rho(x',t) v_i(x',t)}{|x-x'|}\,.\\
             \end{eqnarray}

             The inverse metric of Eq. (\ref{8nl1}) can be calculated using the relation $g^{\alpha \rho} g_{\rho \beta}
            = \delta^\alpha_\beta$, giving the following results:
                \begin{equation}\label{8nl2}
                    \left\{
                    \begin{array}{lll}
                        g^{00} = g^{(0)00} + g^{(2)00} + g^{(4)00} + {\cal O}(6) \ \
                        & \mbox{with} & \ \left\{
                            \begin{array}{l}
                                g^{(0)00} = 1 \\
                                g^{(2)00} = 2 U \\
                                g^{(4)00} = - g_{00}^{(4)} + 4 U^2
                            \end{array}
                        \right. \\
                        \\
                        g^{0i} = g^{(3)0i} + {\cal O}(5) \ \ & \mbox{with} & \ \ g^{(3)0i} = \delta^{ij} g_{0j}^{(3)} \\
                        \\
                        g^{ij} = g^{(0)ij} + g^{(2)ij} + g^{(4)ij} + {\cal O}(6) \ \
                        & \mbox{with} & \ \left\{
                            \begin{array}{l}
                                g^{(0)ij} = - \delta^{ij} \\
                                g^{(2)ij} = \delta^{ij} 2 V \\
                                g^{(4)ij} = - 4 V^2 \delta^{ij} - \delta^{ik} \delta^{jl} g_{kl}^{(4)}
                            \end{array}
                        \right.
                    \end{array}
                    \right.
                \end{equation}

            Given a metric tensor, the associated connection associated can be derived as  
                $\Gamma^\alpha_{\, \mu \nu} = \frac{1}{2} g^{\alpha \beta} \lp \partial_\mu g_{\nu \beta} +
                    \partial_\nu g_{\mu \beta} - \partial_\beta g_{\mu \nu}
                    \rp,$ which can be expanded in powers of $\bar v^2$ introducing, in the last equation, the expressions
            given by Eq. (\ref{8nl1}) and Eq. (\ref{8nl2}), namely:
                \begin{equation}\label{8nl3}
                    \left\{
                    \begin{array}{lll}
                        \Gamma^0_{\, 00} = \Gamma^{(3)0}_{\ \ \ \, 00} + {\cal O}(5) \ \
                        & \mbox{with} & \ \ \Gamma^{(3)0}_{\ \ \ \, 00} = - \partial_0 U \\
                        \\
                        \Gamma^0_{\, 0i} = \Gamma^{(2)0}_{\ \ \ \, 0i} + \Gamma^{(4)0}_{\ \ \ \, 0i} + {\cal O}(6) \ \
                        & \mbox{with} & \ \left\{
                            \begin{array}{l}
                                \Gamma^{(2)0}_{\ \ \ \, 0i} = - \partial_i U \\
                                \\
                                \Gamma^{(4)0}_{\ \ \ \, 0i} = \frac{1}{2} \lp \partial_i g_{00}^{(4)} -
                                4 U \partial_i U \rp
                            \end{array}
                        \right.
                        \\
                        \\
                        \Gamma^0_{\, ij} = \Gamma^{(3)0}_{\ \ \ \, ij} + {\cal O}(5) \ \
                        & \mbox{with} & \ \ \Gamma^{(3)0}_{\ \ \ \, ij} =
                        \frac{1}{2} \lp \partial_j g_{0i}^{(3)} + \partial_i g_{0j}^{(3)} + 2 \delta_{ij} \partial_0
                        V \rp
                        \\
                        \\
                        \Gamma^i_{\, 00} = \Gamma^{(2)i}_{\ \ \ \, 00} + \Gamma^{(4)i}_{\ \ \ \, 00} + {\cal O}(6) \ \
                        & \mbox{with} & \ \left\{
                            \begin{array}{l}
                                \Gamma^{(2)i}_{\ \ \ \, 00} = - \delta^{il} \partial_l U \\
                                \\
                                \Gamma^{(4)i}_{\ \ \ \, 00} = \frac{1}{2} \delta^{il} \lp \partial_l g_{00}^{(4)}
                                + 4 V \partial_l U - 2 \partial_0 g_{0l}^{(3)} \rp
                            \end{array}
                        \right.
                        \\
                        \\
                        \Gamma^i_{\, 0j} = \Gamma^{(3)i}_{\ \ \ \, 0j} + {\cal O}(5) \ \
                        & \mbox{with} & \ \ \Gamma^{(3)i}_{\ \ \ \, 0j} = \frac{1}{2} \delta^{il} \lp \partial_l
                        g_{0j}^{(3)} - \partial_j g_{0l}^{(3)} + 2 \delta_{lj} \partial_0 V \rp
                        \\
                        \\
                        \Gamma^i_{\, jk} = \Gamma^{(2)i}_{\ \ \ \, jk} + \Gamma^{(4)i}_{\ \ \ \, jk} + {\cal O}(6) \ \
                        & \mbox{with} & \ \left\{
                            \begin{array}{l}
                                \Gamma^{(2)i}_{\ \ \ \, jk} = \delta^{il} \lp - \delta_{jk} \partial_l V + \delta_{lj}
                                \partial_k V + \delta_{lk} \partial_j V \rp
                                \\
                                \\
                                \Gamma^{(4)i}_{\ \ \ \, jk} = - \frac{1}{2} \delta^{il} \left[ \partial_j g^{(4)}_{kl} +
                                \partial_k g^{(4)}_{jl} - \partial_l g^{(4)}_{jk} \right] -\\
                                \\
                                \ \ \ \ \ \ \ \ \ \ \ \ - 2 V \delta^{il} \left[
                                \delta_{lk} \partial_j V + \delta_{lj} \partial_k V - \delta_{jk} \partial_l V \right]
                            \end{array}
                        \right.
                    \end{array}
                    \right.
                \end{equation}
Given a metric tensor,  the Riemann tensor, the Ricci tensor and the scalar curvature can be immediately calculated by using the
            following expressions
                \begin{equation}\label{8nl3b}
                    R_{\alpha \beta \rho \sigma} = \frac{1}{2} \lp \partial_\sigma \partial_\alpha g_{\beta \rho}
                    - \partial_\sigma \partial_\beta g_{\alpha \rho} - \partial_\rho \partial_\alpha g_{\beta \sigma}
                    + \partial_\rho \partial_\beta g_{\alpha \sigma} \rp + g_{\mu \nu} \lp \Gamma^{\mu}_{\sigma \alpha}
                    \Gamma^{\nu}_{\beta \rho} - \Gamma^{\mu}_{\rho \alpha} \Gamma^{\nu}_{\beta \sigma} \rp\,,$$
                    $$R_{\mu \nu} = \partial_\rho \Gamma^\rho_{\, \mu \nu} - \partial_\mu \Gamma^\rho_{\, \rho \nu}
                    + \Gamma^\rho_{\, \mu \nu} \Gamma^\sigma_{\, \rho \sigma} - \Gamma^\rho_{\, \sigma \nu}
                    \Gamma^\sigma_{\, \rho \mu},$$
                    $$R = g^{\mu \nu} R_{\mu \nu}.
                \end{equation}
            The components of the Riemann tensor that we will need later can be expanded in powers of $\bar v^2$ using
           Eqs. (\ref{8nl1})-(\ref{8nl3b}):
                \begin{equation}\label{8nl3c}
                    \left\{
                    \begin{array}{lll}
                        R_{i0j0} = R_{i0j0}^{(2)} + R_{i0j0}^{(4)} + {\cal O}(6) \ \ & \mbox{with} & \
                        \left\{
                            \begin{array}{lll}
                                R_{i0j0}^{(2)} & = & \partial_i \partial_j U
                                \\
                                \\
                                R_{i0j0}^{(4)} & = & \frac{1}{2} \lp \partial_0 \partial_i g^{(3)}_{0j} + \partial_0
                                \partial_j g^{(3)}_{0i} + 2 \delta_{ij} \partial_0 \partial_0 V - \partial_i \partial_j
                                g^{(4)}_{00} \rp +
                                \\
                                \\
                                & & + \lp \partial_i U \rp \lp \partial_j U \rp - \lp \partial_i U \rp \lp \partial_j V
                                \rp - \lp \partial_j U \rp \lp \partial_i V \rp +
                                \\
                                \\
                                & & + \delta_{ij} \delta^{kl} \lp \partial_k
                                U \rp \lp \partial_l V \rp
                            \end{array}
                        \right.
                        \\
                        \\
                        R_{0ij0} = R_{0ij0}^{(2)} + R_{0ij0}^{(4)} + {\cal O}(6) \ \ & \mbox{with} & \
                        \left\{
                            \begin{array}{lll}
                                R_{0ij0}^{(2)} & = & - R_{i0j0}^{(2)}
                                \\
                                \\
                                R_{0ij0}^{(4)} & = & - R_{i0j0}^{(4)}
                            \end{array}
                        \right.
                        \\
                        \\
                        R_{ijk0} = R_{ijk0}^{(3)} + {\cal O}(5) \ \ & \mbox{with} & \ \ R_{ijk0}^{(3)} = \frac{1}{2} \left[
                        \partial_k \lp \partial_j g^{(3)}_{0i} - \partial_i g^{(3)}_{0j} \rp + 2
                        \partial_0 \lp \delta_{ik} \partial_j V - \delta_{jk} \partial_i V \rp \right]
                    \end{array}
                    \right.
                \end{equation}
            By assuming the harmonic gauge, given by $g^{\mu \nu} \Gamma^\rho_{\mu \nu} = 0$ (in order to simplify the
            expressions), and using Eqs. (\ref{8nl1})-(\ref{8nl3b}), we can expand the Ricci tensor in powers of $\bar v^2$:
                \begin{equation}\label{8nl4}
                    \left\{
                    \begin{array}{lll}
                        R_{00} = R^{(2)}_{00} + R^{(4)}_{00} + {\cal O}(6) \ \
                        & \mbox{with} & \ \left\{
                            \begin{array}{l}
                                R^{(2)}_{00} = - \triangle U
                                \\
                                \\
                                R^{(4)}_{00} = \frac{1}{2} \triangle g_{00}^{(4)} + 2 V \triangle U + \partial_0
                                \partial_0 U - 2 \delta^{mn} \lp \partial_m U \rp \lp \partial_n U \rp
                            \end{array}
                        \right.
                        \\
                        \\
                        R_{0i} = R_{0i}^{(3)} + {\cal O}(5) \ \ & \mbox{with} & \ \ R_{0i}^{(3)} = \frac{1}{2}
                        \triangle g_{0i}^{(3)}
                        \\
                        \\
                        R_{ij} = R_{ij}^{(2)} + R_{ij}^{(4)} + {\cal O}(6) \ \ & \mbox{with} & \ \ \left\{
                            \begin{array}{lll}
                                R_{ij}^{(2)} & = & - \delta_{ij} \triangle V
                                \\
                                \\
                                R^{(4)}_{ij} & = & \frac{1}{2} \left[ \partial_0 \lp \partial_j g^{(3)}_{0i} + \partial_i
                                g^{(3)}_{0j} + 2 \delta_{ij} \partial_0 V \rp -\right.
                                \\
                                \\
                                & & \left. - \delta^{mn} \lp \partial_m \lp \partial_i
                                g^{(4)}_{nj} + \partial_j g^{(4)}_{ni} \rp - \partial_i \partial_j g^{(4)}_{mn} \rp
                                + \triangle g^{(4)}_{ij} - \partial_i \partial_j g^{(4)}_{00}
                                \right] +
                                \\
                                \\
                                & & + 2 V
                                \partial_i \partial_j V + 2 U \partial_i \partial_j U + \partial_i U \, \partial_j
                                ( U - V ) +
                                \\
                                \\
                                & & + \partial_i V \, \partial_j ( 3 V - U ) + \delta_{ij} \lp \delta^{kl} \partial_l V
                                \partial_k ( U + V ) + 2 V \triangle V \rp
                            \end{array}
                        \right.
                    \end{array}
                    \right.
                \end{equation}
            and the scalar curvature too:
                \begin{equation}\label{8nl5}
                    R = R^{(2)} + R^{(4)} + {\cal O}(6) \ \ \mbox{with} \ \ \left\{
                        \begin{array}{l}
                            R^{(2)} = 3 \triangle V - \triangle U
                            \\
                            \\
                            R^{(4)} = \frac{1}{2} \triangle g_{00}^{(4)} + 2 V \lp \triangle U - 3 \triangle V
                            \rp + \partial_0 \partial_0 U - 2 \delta^{ij} \lp \partial_i U \rp \lp
                            \partial_j U \rp - 2 U \triangle U
                        \end{array}
                    \right.
                \end{equation}

            On the matter side, we start with the general definition of the energy-momentum tensor of a perfect fluid:
                \begin{equation}\label{8nl6}
                    T_{\mu \nu} = \lp \rho + \Pi \rho + p \rp u_\mu u_\nu - p g_{\mu \nu}\,,
                \end{equation}
            where $\Pi$ denotes the internal energy density, $\rho$ the energy density and $p$ the
            pressure. Taking into account that:
                \begin{equation}\label{8nl7}
                    u^0 = \frac{1}{\sqrt{1 - \textbf{v}^2}} = 1 + \frac{1}{2} \textbf{v}^2 + \frac{3}{8} \textbf{v}^4 +
                    {\cal O}(6)\,,$$
                    $$ u^i = \frac{v^i}{\sqrt{1 - \textbf{v}^2}} = v^i \lp 1 + \frac{1}{2} \textbf{v}^2 +
                    {\cal O}(4) \rp\,,
                \end{equation}
            and the expressions Eq. (\ref{8nl1}) and Eq. (\ref{8nl2}), we can calculate the different components of Eq. (\ref{8nl6}):
                \begin{equation}\label{8nl8}
                    \left\{
                    \begin{array}{lll}
                        T_{00} = T_{00}^{(0)} + T_{00}^{(2)} + T_{00}^{(4)} + {\cal O}(6) \ \ & \mbox{with} & \ \left\{
                            \begin{array}{lll}
                                T_{00}^{(0)} & = & \rho
                                \\
                                \\
                                T_{00}^{(2)} & = & \rho \lp \Pi + \textbf{v}^2 - 4 U \rp
                                \\
                                \\
                                T_{00}^{(4)} & = & \rho \left[ \textbf{v}^4 - 3 U \textbf{v}^2 + 4 U^2 + 2 g_{00}^{(4)} + 2
                                g_{0i}^{(3)} v^i  + \right.
                                \\
                                \\
                                & & + \left. \Pi ( \textbf{v}^2 - 4 U ) \right] + p \lp \textbf{v}^2 - 2 U \rp
                            \end{array}
                        \right.
                        \\
                        \\
                        T_{0i} = T_{0i}^{(1)} + T_{0i}^{(3)} + {\cal O}(5) \ \ & \mbox{with} & \ \left\{
                            \begin{array}{l}
                                T_{0i}^{(1)} = - v^i \rho
                                \\
                                \\
                                T_{0i}^{(3)} = - v^i \rho \lp \Pi + \frac{1}{2} \textbf{v}^2 - 2 U +
                                \frac{p}{\rho} \rp
                            \end{array}
                        \right.
                        \\
                        \\
                        T_{ij} = T_{ij}^{(2)} + {\cal O}(4) \ \ & \mbox{with} & \ \ \left\{
                            \begin{array}{lll}
                                T_{ij}^{(2)} & = & \rho \delta_{ik} \delta_{jl} v^k v^l + p \delta_{ij}
                                \\
                                \\
                                T_{ij}^{(4)} & = & \rho \left\{ - \delta_{ik} v^k g_{0j}^{(3)} - \delta_{jl} v^l
                                g_{0i}^{(3)} + \right.
                                \\
                                \\
                                & & \left.+ \delta_{ik} \delta_{jl} v^k v^l \lp \textbf{v}^2 + \Pi + 4 V \rp \right\} +
                                \\
                                \\
                                & & + p \left\{ \delta_{ik} \delta_{jl} v^k v^l + 2 V \delta_{ij} \right\}
                            \end{array}
                        \right.
                    \end{array}
                    \right.
                \end{equation}
   Finally, the expansion of the Gauss-Bonnet invariant $\mathcal G$ in orders of $\bar v^2$ can be calculated
            using Eqs. (\ref{8nl1})-(\ref{8nl5}):
                \begin{equation}\label{8nl9}
                    \mathcal G = \mathcal G^{(4)} + {\cal O}(6)\,,$$
                    $$\mbox{with} \ \ \mathcal G^{(4)} = - 3 \lp \triangle U \rp^2 + \lp \triangle V \rp^2
                    - 6 \lp \triangle U \rp \lp \triangle V \rp + 4 \delta^{im} \delta^{jn} \left[
                    \partial_i \partial_j \lp U + V \rp \right] \left[ \partial_m \partial_n \lp U + V \rp \right]\,.
                \end{equation}

            We have now all the ingredients, expanded in orders of $\bar v^2$, needed to write the field equations and the
            trace equation in the Newtonian, PN and PPN limits. In order to do this, we assume that:
                \begin{equation}\label{8fe1}
                    f^*(R, \mathcal G) = f^*(0,0) + f^*_R (0,0) R + f^*_{\mathcal G}(0,0) \mathcal G + \frac{1}{2} \lp
                    f^*_{RR}(0,0) R^2 + 2 f^*_{R \mathcal G}(0,0) R \mathcal G + f^*_{\mathcal G \mathcal G} \mathcal G^2
                    \rp + ...
                \end{equation}
            where $f^*(R, \mathcal G)$ denotes the function $f(R, \mathcal G)$ or any of its derivatives, i.e. $f_R
            (R, \mathcal G)$, $f_{\mathcal G}(R, \mathcal G)$, $f_{RR}(R, \mathcal G)$ ... Considering (\ref{8nl5}) and
            (\ref{8nl9}), we can write (\ref{8fe1}) as an expansion in orders of $\bar v^2$:
                \begin{equation}\label{8fe2}
                    f^*(R, \mathcal G) = f^*(0,0) + f^*_R(0,0) R^{(2)} + \lp \frac{1}{2} f^*_{RR}(0,0) R^{(2) \, 2} +
                    f^*_R(0,0) R^{(4)} + f^*_{\mathcal G}(0,0) \mathcal G^{(4)} \rp + {\cal O}(6).
                \end{equation}

           In the following sub subsection  we  proceed to calculate the Newtonian, PN and PPN limits.
            % for the first Friedmann equation and the trace equation.

            \subsubsection{The $(0,0)$-field equation}

                The $(0,0)$-field equation is given from Eq. (\ref{8g2}) by:
                    \begin{equation}\label{8fe3}
                        - \frac{1}{2} g_{00} f(R, \mathcal G) + f_R (R, \mathcal G) R_{00} + g_{00} \nabla^2 \lp f_R (R,
                        \mathcal G) \rp - \nabla_0 \nabla_0 \lp f_R (R, \mathcal G) \rp +$$
                        $$+ 2 f_\mathcal{G}(R, \mathcal G) R R_{00} - 4 f_{\mathcal G}(R, \mathcal G) R_{0 \rho}
                        R_{0}^{\ \, \rho} + 2 f_{\mathcal G}(R, \mathcal G) R_{\alpha \beta \rho 0} R^{\alpha \beta
                        \rho}_{\ \ \ \ 0} - 4 f_{\mathcal G}(R, \mathcal G) R_{0 \rho 0 \sigma} R^{\rho \sigma} +$$
                        $$+ 2 g_{00} R \nabla^2 f_{\mathcal G}(R, \mathcal G) - 4 g_{00} R_{\rho \sigma} \nabla^\rho
                        \nabla^\sigma f_{\mathcal G}(R, \mathcal G) - 2 R \nabla_0 \nabla_0 f_{\mathcal G}(R, \mathcal G)
                        - 4 R_{00} \nabla^2 f_{\mathcal G}(R, \mathcal G) +$$
                        $$+ 8 R_{0 \rho} \nabla^\rho \nabla_0 f_{\mathcal G}(R, \mathcal G) + 4 R_{0 \rho 0 \sigma}
                        \nabla^\rho \nabla^\sigma f_{\mathcal G}(R, \mathcal G) = 2 \kappa^2 T_{00},
                    \end{equation}
                At the lowest order in the velocity, we obtain: $f(0,0) = 0$.

                In the Newtonian limit, {\it i.e.} at ${\cal O}(2)$ order in the velocity, Eq.(\ref{8fe3}) reduces to:
                    \begin{equation}\label{8fe4}
                        - \frac{1}{2} g_{00}^{(0)} f_R(0,0) R^{(2)} + f_R (0,0) R_{00}^{(2)} + \left[ g_{00} \nabla^2 \lp
                        f_R (R, \mathcal G) \rp - \nabla_0 \nabla_0 \lp f_R (R, \mathcal G) \rp \right]^{(2)} = 2 \kappa^2
                        T_{00}^{(0)},
                    \end{equation}
                where it has been considered that $f(0,0) = 0$. Introducing Eqs. (\ref{8nl1}), (\ref{8nl4}), (\ref{8nl5}) and
              Eq.  (\ref{8nl8}) into Eq. (\ref{8fe4}), we finally obtain the following equation:
                    \begin{equation}\label{8fe5}
                        f_R(0,0) \lp 3 \triangle V + \triangle U \rp + 2 f_{RR}(0,0) \lp 3 \triangle^2 V -
                        \triangle^2 U \rp = - 4 \kappa^2 \rho.
                    \end{equation}
                Where the notation: $\triangle^2 := \triangle \cdot \triangle$, has been introduced, being $\triangle =
                \delta^{ij} \frac{\partial}{\partial x^i} \frac{\partial}{\partial x^j}$.

                In the PN limit, {\it i.e.} at ${\cal O}(4)$ order in the velocity, Eq.(\ref{8fe3}) reduces to:
                    \begin{equation}\label{8fe6}
\begin{array}{lcl}
                        & f_R(0,0) & \left\{ - \frac{1}{2} g_{00}^{(2)} R^{(2)} -
                        \frac{1}{2} g_{00}^{(0)} R^{(4)} + R_{00}^{(4)} \right\} +
                        \\
                        \\
                        + & f_{\mathcal G}(0,0) & \left\{ - \frac{1}{2} \mathcal G^{(4)} + 2 R^{(2)} R_{00}^{(2)} - 4
                        R_{00}^{(2) \, 2} + 2 \lp R_{\alpha \beta \rho 0} R^{\alpha \beta \rho}_{ \ \ \ \ 0} \rp^{(4)} - 4
                        \lp R_{0 \rho 0 \sigma} R^{\rho \sigma} \rp^{(4)}
                        \right\}+
                        \\
                        \\
                        + & f_{RR}(0,0) & \left\{ - \frac{1}{4} R^{(2) \, 2} + R^{(2)} R_{00}^{(2)} + \lp g_{00}^{(2)}
                        g^{(0) \, ij} + g_{00}^{(0)} g^{(2) \, ij} \rp \partial_i \partial_j R^{(2)} + g^{(0) \, ij} \lp
                        \partial_i \partial_j R^{(4)} - \Gamma^{(2) \, k}_{\ \ \ \, ij} \partial_k R^{(2)} \rp \right\} +
                        \\
                        \\
                        + & f_{R \mathcal G}(0,0) & \left\{ g^{(0) \, ij} \partial_i \partial_j \mathcal G^{(4)} + 2
                        \partial_i \partial_j R^{(2)} \left[ g^{(0) \, ij} \lp R^{(2)} - 2 R_{00}^{(2)} \rp + 2 g^{(0) \,
                        ik} g^{(0) \, jl} \lp - R^{(2)}_{kl} + R^{(2)}_{0k0l} \rp \right]
                        \right\}+
                        \\
                        \\
                        + & f_{RRR}(0,0) & g^{(0) \, ij} \left\{ \partial_i R^{(2)} \partial_j R^{(2)} + R^{(2)}
                        \partial_i \partial_j R^{(2)} \right\} = 2 \kappa^2
                        T_{00}^{(2)},
                        \end{array}
                    \end{equation}
                where it has been considered, once more, that $f(0,0) = 0$. The calculations that make possible to derive
                Eq.(\ref{8fe6}) from Eq.(\ref{8fe3}) are written in Appendix A. Introducing in Eq. (\ref{8fe6}) the results
                obtained in the previous section, we finally obtain:
                    \begin{equation}\label{8fe7}
                        \begin{array}{lcl}
                        & f_R(0,0) & \left\{ \frac{1}{4} \triangle g_{00}^{(4)} + 3 ( U + V ) \triangle V + V \triangle U +
                        \frac{1}{2} \partial_0 \partial_0 U - \delta^{ij} \partial_i U \partial_j U \right\} +
                        \\
                        \\
                        + & f_{RR}(0,0) & \left\{ - \frac{1}{2} \triangle^2 g_{00}^{(4)} + \frac{15}{4} \lp \triangle V
                        \rp^2 - \frac{7}{2} \triangle U \triangle V + \frac{11}{4} \lp \triangle U \rp^2 + 6 ( U + 2 V )
                        \triangle^2 V - 4 V \triangle^2 U - \right.
                        \\
                        \\
                        & & \left. - \partial_0 \partial_0 \triangle U + \delta^{ij} \left[ 3 \partial_i V \, \partial_j \lp 3 \triangle V - \triangle U \rp + 8
                        \partial_i U \, \partial_j \triangle U + 4 \delta^{mn} \partial_i \partial_m U \, \partial_j
                        \partial_n U \right] \right\} +
                        \\
                        \\
                        + & 2 f_{R \mathcal G}(0,0) & \left\{ 4 \lp \triangle U + \triangle V \rp \triangle^2 U - 4
                        \triangle V \triangle^2 V + \delta^{ij} \left[ 3 \partial_i \triangle U \, \partial_j \lp
                        \triangle U + 2 \triangle V \rp - \partial_i \triangle V \, \partial_j \triangle V \right] +
                        \right.
                        \\
                        \\
\end{array}
                    \end{equation}

                In the PPN limit, {\it i.e.} at ${\cal O}(6)$ order in the velocity, using the results obtained in Appendix B,
                Eq.(\ref{8fe3}) reduces to:
                    \begin{equation*}
                        \begin{array}{lcl}
                            & f_R(0,0) & \left\{ - \frac{1}{2} \left[ g^{(4)}_{00} R^{(2)} + g^{(2)}_{00} R^{(4)} +
                            g^{(0)}_{00} R^{(6)} \right] + R^{(6)}_{00} \right\} +
                            \\
                         \\
                          + & f_{\mathcal G}(0,0) & \left\{ - \frac{1}{2} \left[ g_{00}^{(2)} \mathcal G^{(4)} +
                            g_{00}^{(0)} \mathcal G^{(6)} \right] + 2 \left[ R^{(2)} R^{(4)}_{00} + R^{(4)} R^{(2)}_{00}
                            \right] - \right.
                            \\
                            \\
                            & & \left. -4 \left[ 2 g^{(0) \, 00} R^{(2)}_{00} R^{(4)}_{00} + g^{(2) \, 00} R^{(2) \, 2}_{00}
                            + g^{(0) \, ij} R^{(3)}_{0i} R^{(3)}_{0j} \right] + 2 \lp R_{\alpha \beta \rho 0} R^{\alpha \beta \rho}_{\ \ \ \ 0} \rp^{(6)} - 4
                            \lp R_{0 \rho 0 \sigma} R^{\rho \sigma} \rp^{(6)}
                            \right\} +
                            \\
                            \\
                            + & f_{RR}(0,0) & \left\{ - \frac{1}{2} \left[ \frac{1}{2} g_{00}^{(2)} R^{(2) \, 2} +
                            g_{00}^{(0)} R^{(2)} R^{(4)} \right] + R^{(2)} R_{00}^{(4)} + R^{(4)} R_{00}^{(2)} + \right.
                            \\
                            \\
\end{array}
                    \end{equation*}
                    \begin{equation}\label{8fe7b}
                        \begin{array}{lcl}
                            + & f_{RRR}(0,0) & \left\{ R^{(2)} \left[ R^{(2)} \lp - \frac{1}{12} R^{(2)} + \frac{1}{2}
                            R^{(2)}_{00} \rp - \partial_0 \partial_0 R^{(2)} + \Gamma^{(2) \, i}_{\ \ \ \ 00} \partial_i
                            R^{(2)} \right] - \partial_0 R^{(2)} \, \partial_0 R^{(2)} + \right.
                            \\
                            \\
                            & & + g^{(0)}_{00} \left[ g^{(0) \, 00} \lp \partial_0 R^{(2)} \, \partial_0 R^{(2)} +
                            R^{(2)} \partial_0 \partial_0 R^{(2)} - \Gamma^{(2) \, i}_{\ \ \ \ 00} R^{(2)} \partial_i
                            R^{(2)} \rp + \right.
                            \\
                            \\
                            & & + g^{(0) \, ij} \left( 2 \partial_i R^{(2)} \partial_j R^{(4)} + R^{(2)}
                            \partial_i \partial_j R^{(4)} + \right.
                            \\
                            \\
                            & & \left. \left. + R^{(4)} \partial_i \partial_j R^{(2)} - \Gamma^{(2) \, k}_{\ \ \ \ ij}
                            R^{(2)} \partial_k R^{(2)} \right) + g^{(2) \, ij} \lp \partial_i R^{(2)} \partial_j R^{(2)}
                            + R^{(2)} \partial_i \partial_j R^{(2)} \rp \right] +
                            \\
                            \\
\end{array}
                    \end{equation}
                And introducing in Eq. (\ref{8fe7b}) the results obtained in the previous section, we finally obtain:
                    \begin{equation*}
                        \begin{array}{lcl}
                            & f_R(0,0) & \left\{ \frac{1}{2} \lp R^{(6)}_{00} + \delta^{ij} R^{(6)}_{ij} \rp - \delta^{ij}
                            g^{(3)}_{0i} R^{(3)}_{0j} - \frac{1}{2} \triangle V \left[ 3 \lp g^{(4)}_{00} + 4 U V - 4 V^2
                            \rp + \delta^{ij} g^{(4)}_{ij} \right] - \right.
                            \\
                            \\
                            & & - ( U + V ) \left[ \frac{1}{2} \lp 6 \partial_0 \partial_0 V - \triangle g^{(4)}_{00} + 2
                            \delta^{ij} \lp \triangle g^{(4)}_{ij} + \partial_0 \partial_i g^{(3)}_{0j} \rp - 2 \delta^{ij}
                            \delta^{mn} \partial_i \partial_m g^{(4)}_{jn}  \rp \right] +
                            \\
                            \\
                            & & \left. + 8 V \triangle V + 2 U \triangle
                            U + \delta^{ij} \left[ \partial_i U \partial_j ( U - V ) + 2 \partial_i V \partial_j
                            ( 3 V + U ) \right] \right\} +
                            \\
                            \\
                            + & f_{\mathcal G}(0,0) & \left\{ - \frac{1}{2} \mathcal G^{(6)} + ( 8 V - 7 U ) \lp \triangle
                            U \rp^2 + U \lp \triangle V \rp^2 + 2 ( 4 V - 3 U ) \triangle U \triangle V + \right.
                            \\
                            \\
\end{array}
                    \end{equation*}
                    \begin{equation*}
                        \begin{array}{lcl}
                            & & + 2 \triangle U
                            \partial_i U \partial_j V - \frac{1}{2} \partial_i V \partial_j \triangle g_{00}^{(4)} + 2 g_{0i}^{(3)}
                            \partial_0 \partial_j \lp 3 \triangle V - \triangle U \rp + 2 \partial_i \triangle U \lp U
                            \partial_j \lp V - 8 U \rp - 8 V \partial_j U \rp -
                            \\
                            \\
                            & & \left. - 2 \lp V + 3 U \rp \partial_i V \partial_j \lp 3 \triangle V - \triangle U \rp
                            \right] +
                            \\
                            \\
                            & & + \delta^{ij} \delta^{mn} \left[ - \frac{1}{2} \lp 2 \partial_i g_{mj}^{(4)} -
                            \partial_m g_{ij}^{(4)} \rp \partial_n \lp 3 \triangle V - \triangle U \rp - g_{im}^{(4)}
                            \partial_j \partial_n \lp 3 \triangle V - \triangle U \rp - \right.
                            \\
                            \\
                            & & \left. \left. - 4 \lp 2 ( U + V ) \partial_i
                            \partial_m U - \partial_i V \partial_m U \rp \partial_j \partial_n U \right] \right\} +
                            \\
                            \\
\end{array}
                    \end{equation*}
                    \begin{equation}\label{8fe7c}
                        \begin{array}{lcl}
                            & & \left. \left. 2 \partial_j \lp \frac{1}{2} \triangle g_{00}^{(4)} + \partial_0 \partial_0
                            U - 2 U \triangle U \rp \right) + 8 \lp 3 \triangle V - \triangle U \rp \partial_i U
                            \partial_j \triangle U \right] +
                            \\
                            \\
                            & & \left. + 4 \delta^{ij} \delta^{mn} \left[ \partial_i \partial_m \lp 3 \triangle V -
                            \triangle U \rp \partial_j \partial_n \lp 3 \triangle V - \triangle U \rp + \lp 3 \triangle V
                            - \triangle U \rp \partial_i \partial_m U \partial_j \partial_n U \right] \right\} +
                            \\
                            \\
                            + & f_{RR \mathcal G}(0,0) & \left\{ \triangle^2 U \left[ - 11 \lp \triangle U \rp^2 + 25 \lp
                            \triangle V \rp^2 + 10 \triangle U \triangle V + 4 \delta^{ij} \delta^{mn} \partial_i
                            \partial_m ( U + V ) \partial_j \partial_n ( U + V ) \right] + \right.
                            \\
                            \\
                            & & + \triangle^2 V \left[ 9 \lp \triangle U \rp^2 - 39 \lp \triangle V \rp^2 + 26 \triangle U
                            \triangle V - 12 \delta^{ij} \delta^{mn} \partial_i \partial_m ( U + V ) \partial_j \partial_n
                            ( U + V ) \right] +
                            \\
                            \\
\end{array}
                    \end{equation}

                Then, Eq.(\ref{8fe5}), Eq.(\ref{8fe7}) and Eq.(\ref{8fe7c}) constitute the Newtonian, PN and PPN limits,
                respectively, for the $(0,0)$-field equation of $f(R, \mathcal G)$-gravity when the metric tensor
               Eq. (\ref{8nl1}) is assumed.

            \subsubsection{The trace equation}

                We proceed to find the Newtonian, PN and PPN limits for the trace equation. From Eq.(\ref{8g3}) at
                ${\cal O}(0)$ order in the velocity we obtain again: $f(0,0) = 0$.

                In the Newtonian limit, {\it i.e.} at ${\cal O}(2)$ order in the velocity, Eq.(\ref{8g3}) reduces to:
                    \begin{equation}\label{8fe8}
                        - f_R(0,0) R^{(2)} + 3 f_{RR}(0,0) \lp \nabla^2 R \rp^{(2)} = 2 \kappa^2 T^{(0)}.
                    \end{equation}
                Using Eq. (\ref{8nl2}), (\ref{8nl5}) and  Eq.(\ref{8nl8}), Eq.(\ref{8fe8}) is given by:
                    \begin{equation}\label{8fe9}
                        f_R(0,0) \lp 3 \triangle V - \triangle U \rp + 3 f_{RR}(0,0) \lp 3 \triangle^2 V - \triangle^2 U
                        \rp = - 2 \kappa^2 \rho.
                    \end{equation}

                In the PN limit, {\it i.e.} at ${\cal O}(4)$ order in the velocity, using the calculations given in Appendix A,
                Eq.(\ref{8g3}) reduces to:
                    \begin{equation}\label{8fe10}
\begin{array}{lcl}
                        + & 3 f_{RR}(0,0) & \left\{ g^{(0) \, 00} \lp \partial_0 \partial_0 R^{(2)} -
                        \Gamma^{(2) \, i}_{\ \ \ \ 00} \partial_i R^{(2)} \rp + g^{(0) \, ij} \lp \partial_i \partial_j
                        R^{(4)} - \Gamma^{(2) \, k}_{\ \ \ \, ij} \partial_k R^{(2)} \rp + g^{(2) \, ij} \partial_i
                        \partial_j R^{(2)} \right\} +
                        \\
                        \\
                        + & 3 f_{RRR}(0,0) & g^{(0) \, ij} \left\{ \partial_i R^{(2)} \partial_j R^{(2)} + R^{(2)}
                        \partial_i \partial_j R^{(2)} \right\} +
                        \\
                        \\
                        + & f_{R \mathcal G}(0,0) & \left\{ 3 g^{(0) \, ij} \partial_i \partial_j \mathcal G^{(4)} + 2
                        R^{(2)} g^{(0) \, ij} \partial_i \partial_j R^{(2)} - 4 g^{(0) \, im} g^{(0) \, jn} R^{(2)}_{ij}
                        \partial_m \partial_n R^{(2)} \right\} -
                        \\
                        \\
                        - & f_R(0,0) R^{(4)} & =  2 \kappa^2 \left\{ g^{(0) \, 00} T^{(2)}_{00} + g^{(2) \, 00} T^{(0)}_{00} + g^{(0) \, ij}
                        T^{(2)}_{ij} \right\}.
                        \end{array}
                    \end{equation}
                Using in Eq. (\ref{8fe10}) the results obtained in the previous section, the PN limit for the trace equation is
                given by:
                    \begin{equation}\label{8fe11}
                        \begin{array}{lcl}
                        - & f_R(0,0) & \left\{ \frac{1}{2} \triangle g_{00}^{(4)} + 2 V \lp \triangle U - 3 \triangle V
                        \rp + \partial_0 \partial_0 U - 2 \delta^{ij} \lp \partial_i U \rp \lp \partial_j U \rp - 2 U
                        \triangle U \right\} +
                        \\
                        \\
                        + & 3 f_{RR}(0,0) & \left\{ - \frac{1}{2} \triangle^2 g_{00}^{(4)} + 2 \lp \triangle U \rp^2 - 2
                        \triangle U \, \triangle V + 6 \lp \triangle V \rp^2 + \partial_0 \partial_0 \lp 3 \triangle V -
                        2 \triangle U \rp + 2 \lp U - 2 V \rp \triangle^2 U + \right.
                        \\
                        \\
                        & & \left. + 12 V \triangle^2 V + \delta^{ij} \left[ \partial_i \lp 3 V + U \rp \partial_j \lp 3 \triangle V -
                        \triangle U \rp + 8 \partial_i U \, \partial_j \triangle U + 4 \delta^{mn} \lp \partial_i
                        \partial_m U \rp \lp \partial_j \partial_n U \rp \right] \right\} -
                        \\
                        \\
                        - & 3 f_{RRR}(0,0) & \left\{ \delta^{ij} \partial_i \lp 3 \triangle V - \triangle U \rp
                        \partial_j \lp 3 \triangle V - \triangle U \rp + \lp 3 \triangle V - \triangle U \rp \triangle \lp 3
                        \triangle V - \triangle U \rp \right\} +
                        \\
                        \\
\end{array}
                    \end{equation}

                In the PPN limit, {\it i.e.} at ${\cal O}(6)$ order in the velocity, using the calculations given in Appendix B,
                Eq.(\ref{8g3}) reduces to:
                    \begin{equation*}
                        \begin{array}{lcl}
                            - & f_R (0,0) R^{(6)} & +
                            \\
                            \\
                            + & 3 f_{RR}(0,0) & \left\{ g^{(0) \, 00} \left[ \partial_0 \partial_0 R^{(4)} -
                            \Gamma^{(3) \, 0}_{\ \ \ \ 00} \partial_0 R^{(2)} - \Gamma^{(2) \, i}_{\ \ \ \ 00} \partial_i
                            R^{(4)} - \Gamma^{(4) \, i}_{\ \ \ \ 00} \partial_i R^{(2)} \right]
                            + \right.
                            \\
                            \\
                            & & + g^{(2) \, 00} \left[
                            \partial_0 \partial_0 R^{(2)} - \Gamma^{(2) \, i}_{\ \ \ \ 00} \partial_i R^{(2)} \right]
                            + g^{(0) \,
                            ij} \left[ \partial_i \partial_j R^{(6)} - \Gamma^{(3) \, 0}_{\ \ \ \ ij} \partial_0 R^{(2)} -
                            \Gamma^{(2) \, k}_{\ \ \ \ ij} \partial_k R^{(4)} -
                            \right.
                            \\
                            \\
                            & & \left. \left. - \Gamma^{(4) \, k}_{\ \ \ \ ij} \partial_k
                            R^{(2)} \right] + 2 g^{(3) \, 0i} \partial_0 \partial_i R^{(2)} + g^{(2) \, ij} \left[ \partial_i \partial_j R^{(4)}
                            - \Gamma^{(2) \, k}_{\ \ \ \ ij} \partial_k R^{(2)} \right] + g^{(4) \, ij} \partial_i
                            \partial_j R^{(2)} \right\} +
                            \\
                            \\
\end{array}
                    \end{equation*}
                    \begin{equation}\label{8fe12}
                        \begin{array}{lcl}
                            + & 3 f_{RRRR}(0,0) & g^{(0) \, ij} R^{(2)} \left\{ \partial_i
                            R^{(2)} \partial_j R^{(2)} +
                            \frac{1}{2} R^{(2)} \partial_i \partial_j R^{(2)}
                            \right\} =
                            \\
                            \\
                            = & 2 \kappa^2 & \left\{ g^{(0) \, 00} T^{(4)}_{00} + g^{(2) \, 00} T^{(2)}_{00} +
                            g^{(4) \, 00} T^{(0)}_{00} + 2 g^{(3) \, 0i} T^{(1)}_{0i} + g^{(0) \, ij} T^{(4)}_{ij} +
                            g^{(2) \, ij} T^{(2)}_{ij} \right\}
                        \end{array}
                    \end{equation}
                Introducing in Eq. (\ref{8fe12}) the results obtained in the previous section, we finally obtain:
                    \begin{equation*}
                        \begin{array}{lcl}
                            - & f_R (0,0) & R^{(6)} +
                            \\
                            \\
                            + & f_{RR}(0,0) & 3 \left\{ \partial_0 \partial_0 \left[ \frac{1}{2} \triangle g_{00}^{(4)}
                            - 2 V \lp 3 \triangle V - \triangle U \rp + \partial_0 \partial_0 U - 2 \delta^{ij} \lp
                            \partial_i U \rp \lp \partial_j U \rp - 2 U \triangle U \right] + \right.
                            \\
                            \\
                            & & + 2 U \partial_0 \partial_0
                            \lp 3 \triangle V - \triangle U \rp + \, \partial_0 ( 3 V + U ) \, \partial_0 \lp 3 \triangle V -
                            \triangle U \rp - \triangle R^{(6)} + V \triangle^2
                            g_{00}^{(4)} +
                            \\
                            \\
                            & & + 4 V ( 2 V - U ) \triangle^2
                            U - 24 V \triangle^2 V - 4 V \triangle U \triangle \lp U - V \rp -
                            \\
                            \\
\end{array}
                    \end{equation*}
                    \begin{equation}\label{8fe13}
                        \begin{array}{lcl}
                            + & f_{\mathcal G \mathcal G}(0,0) & 2 \lp \triangle U - \triangle V
                            \rp \triangle \left\{
                            - 3 \lp \triangle U \rp^2 + \lp \triangle V \rp^2 - 6 \lp \triangle U \rp \lp \triangle V \rp
                            + 4 \delta^{im} \delta^{jn} \partial_i \partial_j \lp U + V \rp \, \partial_m \partial_n \lp U
                            + V \rp \right\} +
                            \\
                            \\
                            + & f_{RRR}(0,0) & \left\{ - \frac{3}{2} \lp 3 \triangle V - \triangle U \rp \triangle^2
                            g_{00}^{(4)} - \frac{37}{6} \lp \triangle U \rp^3 + \frac{51}{2} \lp \triangle U \rp^2
                            \triangle V - \frac{81}{2} \triangle U \lp \triangle V \rp^2 + \frac{117}{2} \lp \triangle V
                            \rp^3 + \right.
                            \\
                            \\
                            & & + 3 \triangle^2 U \left[ \frac{1}{2} \triangle g_{00}^{(4)} + \partial_0
                            \partial_0 U - 2 ( U + 3 V ) \lp 3 \triangle V - \triangle U \rp - 2 \delta^{ij} \partial_i U
                            \partial_j U - 2 U \triangle U \right] +
                            \\
                            \\
                            & & + 9 \triangle^2 V \left[ - \frac{1}{2} \triangle g_{00}^{(4)} - \partial_0 \partial_0 U
                            + 6 V \lp 3 \triangle V - \triangle U \rp + 2 \delta^{ij} \partial_i U \partial_j U + 2 U
                            \triangle U \right] +
                            \\
                            \\
\end{array}
                    \end{equation}

                Then, Eq.(\ref{8fe9}), Eq.(\ref{8fe11}) and Eq.(\ref{8fe13}) constitute the Newtonian, PN and PPN limit,
                respectively, for the trace equation of $f(R, \mathcal G)$- gravity when the metric
                tensor given by Eq. (\ref{8nl1}) is assumed.

%%%%%%%%%%%%%%%%%%%%%%%%%%%%%%%%%%%%%
%%%%%%%%%%%%%%%%%%%%%%%%%%%%%%%%%%%%%
\section{Solving the field equations in the Newtonian limit}
%%%%%%%%%%%%%%%%%%%%%%%%%%%%%%%%%%%%%
%%%%%%%%%%%%%%%%%%%%%%%%%%%%%%%%%%%%%

        Our aim is now  to solve the system of equations for the Newtonian limit of $f(R, \mathcal G)$-
        gravity, {\it i.e.} the system constituted by Eq.(\ref{8fe5}) and Eq.(\ref{8fe9}), in the most general way. In order to
        do this, we will search for solutions in terms of Green's functions (see \cite{Capozziello:2010gu}).

        We start by considering the set of equations given by Eq.(\ref{8fe5}) and Eq.(\ref{8fe9}), this is:
            \begin{equation}\label{8snl1}
                \begin{array}{rcl}
                    f_R(0,0) \lp 3 \triangle V + \triangle U \rp + 2 f_{RR}(0,0) \lp 3 \triangle^2 V -
                    \triangle^2 U \rp & = & - 4 \kappa^2 \rho\,,
                    \\
                    \\
                    f_R(0,0) \lp 3 \triangle V - \triangle U \rp + 3 f_{RR}(0,0) \lp 3 \triangle^2 V - \triangle^2 U
                    \rp & = & - 2 \kappa^2 \rho\,.
                \end{array}
            \end{equation}
        By introducing the new auxiliary functions $A = f_R(0,0) ( 3 V + U )$ and $B = 2 f_{RR}(0,0) \lp 3 V - U \rp$, we
        can write Eq. (\ref{8snl1}) as:
            \begin{equation}\label{8snl2}
                \begin{array}{rcl}
                    - 4 \kappa^2 \rho & = & \triangle A + \triangle^2 B\,,
                    \\
                    \\
                    - 4 \kappa^2 \rho & = & \frac{f_R(0,0)}{f_{RR}(0,0)} \triangle B + 3 \triangle^2 B\,.
                \end{array}
            \end{equation}
        Considering now the new function $\Phi = A + \triangle B$, Eq. (\ref{8snl2}) reduces to:
            \begin{equation}\label{8snl3}
                \begin{array}{rcl}
                    - 4 \kappa^2 \rho & = & \triangle \Phi\,,
                    \\
                    \\
                    - 4 \kappa^2 \rho & = & \frac{f_R(0,0)}{f_{RR}(0,0)} \triangle B + 3 \triangle^2 B\,.
                \end{array}
            \end{equation}
        It is important to remark that Eq. (\ref{8snl3}) is a set of uncoupled equations. We are interested in the solution of
        the second equation in Eq. (\ref{8snl3}) in terms of the Green's function $\mathbb{G}(\mbox{{\bf x}},\mbox{{\bf x'}})$
        defined by:
            \begin{equation}\label{8snl4}
                B = - 4 \kappa^2 C \int d^3 \mbox{{\bf x'}} \mathbb{G}(\mbox{{\bf x}},\mbox{{\bf x'}})\,,
                \rho(\mbox{{\bf x'}})
            \end{equation}
        where $C$ is a constant, which is introduced for dimensional reasons. Now the set of equations given by
       Eq.  (\ref{8snl3}) is equivalent to:
            \begin{equation}\label{8snl5}
                \begin{array}{rcl}
                    - 4 \kappa^2 \rho & = & \triangle \Phi\,,
                    \\
                    \\
                    \frac{1}{C}
                    \delta(\mbox{{\bf x}} - \mbox{{\bf x'}}) & = & \frac{f_R(0,0)}{f_{RR}(0,0)} \triangle_{\mbox{{\bf x}}}
                    \mathbb{G}(\mbox{{\bf x}},\mbox{{\bf x'}}) +
                    3 \triangle^2_{\mbox{{\bf x}}} \mathbb{G}(\mbox{{\bf x}},\mbox{{\bf x'}})\,,
                \end{array}
            \end{equation}
        where $\delta(\mbox{{\bf x}} - \mbox{{\bf x'}})$ is the three-dimensional Dirac $\delta$-function. The general
        solutions of Eqs.(\ref{8snl1}) for $U(\mbox{{\bf x}})$ and $V(\mbox{{\bf x}})$, in terms of the Green's
        function $\mathbb{G}(\mbox{{\bf x}},\mbox{{\bf x'}})$ and the function $\Phi(\mbox{{\bf x}})$, are:
            \begin{equation}\label{8snl6}
                \begin{array}{rcl}
                    U(\mbox{{\bf x}}) & = & \frac{\displaystyle 1}{\displaystyle 2 f_R(0,0)} \Phi(\mbox{{\bf x}}) + 2
                    \kappa^2 C \lp \frac{\displaystyle \triangle_{\mbox{{\bf x}}}}{\displaystyle f_R(0,0)} +
                    \frac{\displaystyle 1}{\displaystyle 2 f_{RR}(0,0)} \rp \displaystyle \int d^3 \mbox{{\bf x'}}
                    \mathbb{G}(\mbox{{\bf x}},\mbox{{\bf x'}}) \rho(\mbox{{\bf x'}})\,,
                    \\
                    \\
                    V(\mbox{{\bf x}}) & = & \frac{\displaystyle 1}{\displaystyle 6 f_R(0,0)} \Phi(\mbox{{\bf x}}) +
                    \frac{2}{3} \kappa^2 C \lp \frac{\displaystyle \triangle_{\mbox{{\bf x}}}}{\displaystyle f_R(0,0)} -
                    \frac{\displaystyle 1}{\displaystyle 2 f_{RR}(0,0)} \rp \displaystyle \int d^3 \mbox{{\bf x'}}
                    \mathbb{G}(\mbox{{\bf x}},\mbox{{\bf x'}}) \rho(\mbox{{\bf x'}})\,.
                \end{array}
            \end{equation}

        In summary, the functions $U(\mbox{{\bf x}})$ and $V(\mbox{{\bf x}})$, which are related with $g_{00}^{(2)}$ and
        $g_{ij}^{(2)}$, respectively, by Eq.(\ref{8nl1}) have been found in terms of the
        Green's function $\mathbb{G}(\mbox{{\bf x}},\mbox{{\bf
        x'}})$ and the function $\Phi(\mbox{{\bf x}})$, giving in this way a general solution to the Newtonian limit for $f(R,\mathcal G)$
       -gravity theories.

%%%%%%%%%%%%%%%%%%%%%%%%%%%%%%%%%%%%%%%%%%%%%%%%%%%%%%%%%%%%%%%%%%%%%%%%%%%%%%%%%%%%%%%%%%%%%%%%
%%%%%%%%%%%%%%%%%%%%%%%%%%%%%%%%%%%%%%%%%%%%%%%%%%%%%%%%%%%%%%%%%%%%%%%%%%%%%%%%%%%%%%%%%%%%%%%%
\section{The weak field limit in two special cases: $f(R)$ and $f(\mathcal{G})$  gravities}
%%%%%%%%%%%%%%%%%%%%%%%%%%%%%%%%%%%%%%%%%%%%%%%%%%%%%%%%%%%%%%%%%%%%%%%%%%%%%%%%%%%%%%%%%%%%%%%%
%%%%%%%%%%%%%%%%%%%%%%%%%%%%%%%%%%%%%%%%%%%%%%%%%%%%%%%%%%%%%%%%%%%%%%%%%%%%%%%%%%%%%%%%%%%%%%%%

        The results obtained previously will be used for two special cases: $f(R)$- gravity and
        $f(\mathcal G)$-gravity, respectively.

        \subsection{$f(R)$  gravity}

            The starting action is given by:
                \begin{equation}\label{8fr1}
                    S=\int d^4x\sqrt{-g} \left( \frac{1}{2 \kappa^2} f(R) + \mathcal{L}_m\right).
                \end{equation}

            In order to obtain the Newtonian, PN and PPN limits for this theory we will use the equations of the previous
            section considering the change: $f(R, \mathcal G) \rightarrow f(R)$. The field equations for $f(R)$-
            gravity are obtained from Eq.(\ref{8g2}):
                \begin{equation}\label{8fr2}
                    - \frac{1}{2} g_{\mu \nu} f(R) + f'(R) R_{\mu \nu} + g_{\mu \nu} \nabla^2 f'(R) - \nabla_\mu
                    \nabla_\nu f'(R) = 2 \kappa^2 T_{\mu \nu},
                \end{equation}
            where primes denote the derivative with respect to Ricci scalar,
            while the trace equation is obtained from Eq.(\ref{8g3}):
                \begin{equation}\label{8fr3}
                    - 2 f(R) + f'(R) R + 3 \nabla^2 f'(R) = 2 \kappa^2 T.
                \end{equation}
            Before analyzing the Newtonian, PN and PPN limits for this theory, it is important to remark that at the
            lowest order in the velocity, {\it i.e.} ${\cal O}(0)$-order, from Eq.(\ref{8fr2}) and Eq.(\ref{8fr3}), we obtain: $f(0) = 0$.

            \subsubsection{The Newtonian limit}

                The Newtonian limit of $f(R)$-gravity corresponds to ${\cal O}(2)$-order for Eq.(\ref{8fr2}) and
                Eq.(\ref{8fr3}).

                The $(0,0)$-field equation for $f(R)$- gravity at Newtonian order can be obtained from Eq. (\ref{8fe5})
                and it is given by:
                    \begin{equation}\label{8fr4}
                        f'(0) \lp 3 \triangle V + \triangle U \rp + 2 f''(0) \lp 3 \triangle^2 V - \triangle^2 U \rp
                        = - 4 \kappa^2 \rho.
                    \end{equation}

                The trace equation for $f(R)$ modified gravity at Newtonian order can be obtained from Eq.(\ref{8fe9}) and
                it is given by:
                    \begin{equation}\label{8fr5}
                        f'(0) \lp 3 \triangle V - \triangle U \rp + 3 f''(0) \lp 3 \triangle^2 V - \triangle^2 U \rp
                        = - 2 \kappa^2 \rho.
                    \end{equation}

            \subsubsection{The Post Newtonian limit}

                In this case, the aim is to obtain Eq.(\ref{8fr2}) and Eq.(\ref{8fr3}) at ${\cal O}(4)$-order in the velocity.

                For the $(0,0)$-field equation, we use Eq.(\ref{8fe7}). For $f(R)$-gravity it reduces to:
                    \begin{equation}\label{8fr4b}
                        \begin{array}{lcl}
                            & f'(0) & \left\{ \frac{1}{4} \triangle g_{00}^{(4)} + 3 ( U + V ) \triangle V + V \triangle U +
                            \frac{1}{2} \partial_0 \partial_0 U - \delta^{ij} \partial_i U \partial_j U \right\} +
                            \\
                            \\
                            + & f''(0) & \left\{ - \frac{1}{2} \triangle^2 g_{00}^{(4)} + \frac{15}{4} \lp \triangle V
                            \rp^2 - \frac{7}{2} \triangle U \triangle V + \frac{11}{4} \lp \triangle U \rp^2 + 6 ( U + 2 V )
                            \triangle^2 V - 4 V \triangle^2 U - \partial_0 \partial_0 \triangle U + \right.
                            \\
                            \\
                            & & + \left. \delta^{ij} \left[ 3 \partial_i V \, \partial_j \lp 3 \triangle V - \triangle U \rp + 8
                            \partial_i U \, \partial_j \triangle U + 4 \delta^{mn} \partial_i \partial_m U \, \partial_j
                            \partial_n U \right] \right\} -
                            \\
                            \\
                            - & f'''(0) & \left\{ \delta^{ij} \partial_i \lp 3 \triangle V - \triangle U \rp \partial_j
                            \lp 3 \triangle V - \triangle U \rp + \lp 3 \triangle V - \triangle U \rp \triangle \lp 3
                            \triangle V - \triangle U \rp \right\} = 2 \kappa^2 \rho \lp \Pi + \textbf{v}^2 - 4 U \rp.
                        \end{array}
                    \end{equation}
                While for the trace equation, Eq.(\ref{8fe11}) gives:
                    \begin{equation}\label{8fr6}
                        \begin{array}{lll}
                        - & f'(0) & \left\{ \frac{1}{2} \triangle g_{00}^{(4)} + 2 V \lp \triangle U - 3 \triangle V
                        \rp + \partial_0 \partial_0 U - 2 \delta^{ij} \lp \partial_i U \rp \lp \partial_j U \rp - 2 U
                        \triangle U \right\} +
                        \\
                        \\
                        + & 3 f''(0) & \left\{ - \frac{1}{2} \triangle^2 g_{00}^{(4)} + 2 \lp \triangle U \rp^2 - 2
                        \triangle U \, \triangle V + 6 \lp \triangle V \rp^2 + \partial_0 \partial_0 \lp 3 \triangle V -
                        2 \triangle U \rp + 2 \lp U - 2 V \rp \triangle^2 U + \right.
                        \\
                        \\
                        & & \left. + 12 V \triangle^2 V + \delta^{ij} \left[ \partial_i \lp 3 V + U \rp \partial_j \lp 3 \triangle V -
                        \triangle U \rp + 8 \partial_i U \, \partial_j \triangle U + 4 \delta^{mn} \lp \partial_i
                        \partial_m U \rp \lp \partial_j \partial_n U \rp \right] \right\} -
                        \\
                        \\
                        - & 3 f'''(0) & \left\{ \delta^{ij} \partial_i \lp 3 \triangle V - \triangle U \rp
                        \partial_j \lp 3 \triangle V - \triangle U \rp + \lp 3 \triangle V - \triangle U \rp \triangle \lp 3
                        \triangle V - \triangle U \rp \right\} = 2 \kappa^2 \left\{ \rho \lp \Pi - 2 U \rp - 3 p \right\}
                        \end{array}
                    \end{equation}

            \subsubsection{The Parameterized Post Newtonian limit}

                Finally, the PPN limit corresponds to ${\cal O}(6)$-order for Eq.(\ref{8fr2}) and Eq.(\ref{8fr3}).

                For the $(0,0)$-field equation we obtain from Eq.(\ref{8fe7c}) the following result:
                    \begin{equation*}
                        \begin{array}{lcl}
                            & f'(0) & \left\{ \frac{1}{2} \lp R^{(6)}_{00} + \delta^{ij} R^{(6)}_{ij} \rp - \delta^{ij}
                            g^{(3)}_{0i} R^{(3)}_{0j} - \frac{1}{2} \triangle V \left[ 3 \lp g^{(4)}_{00} + 4 U V - 4 V^2
                            \rp + \delta^{ij} g^{(4)}_{ij} \right] - \right.
                            \\
                            \\
                            & & - ( U + V ) \left[ \frac{1}{2} \lp 6 \partial_0 \partial_0 V - \triangle g^{(4)}_{00} + 2
                            \delta^{ij} \lp \triangle g^{(4)}_{ij} + \partial_0 \partial_i g^{(3)}_{0j} \rp - 2 \delta^{ij}
                            \delta^{mn} \partial_i \partial_m g^{(4)}_{jn}  \rp \right] +
                            \\
                            \\
                            & & \left. + 8 V \triangle V + 2 U \triangle
                            U + \delta^{ij} \left[ \partial_i U \partial_j ( U - V ) + 2 \partial_i V \partial_j
                            ( 3 V + U ) \right] \right\} +
                            \\
                            \\
                            + & f''(0) & \left\{ - \triangle R^{(6)} + ( U + V ) \triangle^2 g_{00}^{(4)} +
                            \triangle^2 U \lp g_{00}^{(4)} - 4 U^2 + 4 U V + 8 V^2 \rp - 3 \triangle^2 V \lp g_{00}^{(4)}
                            + 4 U V + 8 V^2 \rp - \right.
                            \\
                            \\
                            & & - \lp \triangle U \rp^2 \lp \frac{5}{2} U + 7 V \rp - 3 \lp \triangle V \rp^2 \lp
                            \frac{5}{2} U + V \rp + 2 \triangle U \triangle V \lp 2 U + 5 V \rp +
                            \\
                            \\
                            & & + 3 \lp \triangle V - \triangle U \rp \lp \frac{1}{4} \triangle g_{00}^{(4)} + \frac{1}{2}
                            \partial_0 \partial_0 U - \delta^{ij} \partial_i U \partial_j U \rp + 2 V \partial_0
                            \partial_0 \triangle U +
                            \\
                            \\
                            & & + \delta^{ij} \left[ \frac{1}{2} \lp 2 \partial_i g_{0j}^{(3)} - \partial_0 g_{ij}^{(2)}
                            \rp \partial_0 \lp 3 \triangle V - \triangle U \rp - \partial_i V \partial_0 \partial_0
                            \partial_j U + 2 \lp 3 \triangle V - \triangle U \rp \partial_i V \partial_j V + \right.
                            \\
                            \\
                            & & + 2 \triangle U
                            \partial_i U \partial_j V - \frac{1}{2} \partial_i V \partial_j \triangle g_{00}^{(4)} + 2 g_{0i}^{(3)}
                            \partial_0 \partial_j \lp 3 \triangle V - \triangle U \rp + 2 \partial_i \triangle U \lp U
                            \partial_j \lp V - 8 U \rp - 8 V \partial_j U \rp -
                        \end{array}
                    \end{equation*}
                    \begin{equation}\label{8fr4c}
                        \begin{array}{lcl}
                            & & \left. - 2 \lp V + 3 U \rp \partial_i V \partial_j \lp 3 \triangle V - \triangle U \rp
                            \right] + \delta^{ij} \delta^{mn} \left[ - \frac{1}{2} \lp 2 \partial_i g_{mj}^{(4)} -
                            \partial_m g_{ij}^{(4)} \rp \partial_n \lp 3 \triangle V - \triangle U \rp - \right.
                            \\
                            \\
                            & & \left. \left. - g_{im}^{(4)}
                            \partial_j \partial_n \lp 3 \triangle V - \triangle U \rp - 4 \lp 2 ( U + V ) \partial_i
                            \partial_m U - \partial_i V \partial_m U \rp \partial_j \partial_n U \right] \right\} +
                            \\
                            \\
                            + & f'''(0) & \left\{ \triangle^2 U \left[ \frac{1}{2} \triangle g_{00}^{(4)} +
                            \partial_0 \partial_0 U + 2 ( 3 V - U ) \triangle U - 18 V \triangle V - 2 \delta^{ij}
                            \partial_i U \partial_j U \right] + \right.
                            \\
                            \\
                            & & + \triangle^2 V \left[ - \frac{1}{2} \triangle g_{00}^{(4)} - \partial_0 \partial_0
                            U - 2 ( 7 V + 2 U ) \triangle U + 6 ( 7 V + 3 U ) \triangle V + 2 \delta^{ij} \partial_i U
                            \partial_j U \right] -
                            \\
                            \\
                            & & - \lp \frac{1}{2} \triangle g_{00}^{(4)} + \partial_0 \partial_0 \triangle U \rp
                            \lp 3 \triangle V - \triangle U \rp - \frac{29}{12} \lp \triangle U \rp^3 + \frac{63}{4} \lp
                            \triangle V \rp^3 + \frac{111}{12} \triangle V \lp \triangle U \rp^2 - \frac{57}{4} \triangle
                            U \lp \triangle V \rp^2 +
                            \\
                            \\
                            & & + \delta^{ij} \left[ \partial_i \lp 3 \triangle V - \triangle U \rp \left( 3 \lp 3
                            \triangle V - \triangle U \rp \partial_j V + 2 \lp 3 \triangle V + \triangle U \rp \partial_j
                            \lp 3 \triangle V - \triangle U \rp - \right. \right.
                            \\
                            \\
                            & & \left. \left. 2 \partial_j \lp \frac{1}{2} \triangle g_{00}^{(4)} + \partial_0 \partial_0
                            U - 2 U \triangle U \rp \right) + 8 \lp 3 \triangle V - \triangle U \rp \partial_i U
                            \partial_j \triangle U \right] +
                            \\
                            \\
                            & & \left. + 4 \delta^{ij} \delta^{mn} \left[ \partial_i \partial_m \lp 3 \triangle V -
                            \triangle U \rp \partial_j \partial_n \lp 3 \triangle V - \triangle U \rp + \lp 3 \triangle V
                            - \triangle U \rp \partial_i \partial_m U \partial_j \partial_n U \right] \right\} +
                            \\
                            \\
                            + & f''''(0) & \left\{ - \frac{1}{2} \lp 3 \triangle V - \triangle U \rp^2 \lp 3
                            \triangle^2 V - \triangle^2 U \rp - \delta^{ij} \lp 3 \triangle V - \triangle U \rp \partial_i
                            \lp 3 \triangle V - \triangle U \rp \partial_j \lp 3 \triangle V - \triangle U \rp \right\}
                            =
                            \\
                            \\
                            = & 2 \kappa^2 & \left\{ \rho
                            \left[ \textbf{v}^4 - 3 U \textbf{v}^2 + 4 U^2 + 2 g_{00}^{(4)} + 2 g_{0i}^{(3)} v^i + \Pi (
                            \textbf{v}^2 - 4 U ) \right] + p \lp \textbf{v}^2 - 2 U \rp \right\}
                        \end{array}
                    \end{equation}
                And Eq.(\ref{8fe13}) implies that the trace equation is given by:
                    \begin{equation*}
                        \begin{array}{llcl}
                            & - & f'(0) R^{(6)} & +
                            \\
                            \\
                            & + & 3 f''(0) & \left\{ \partial_0 \partial_0 \left[ \frac{1}{2} \triangle g_{00}^{(4)}
                            - 2 V \lp 3 \triangle V - \triangle U \rp + \partial_0 \partial_0 U - 2 \delta^{ij} \lp
                            \partial_i U \rp \lp \partial_j U \rp - 2 U \triangle U \right] + \right.
                            \\
                            \\
                            & & & + 2 U \partial_0 \partial_0
                            \lp 3 \triangle V - \triangle U \rp + \, \partial_0 ( 3 V + U ) \, \partial_0 \lp 3 \triangle V -
                            \triangle U \rp - \triangle R^{(6)} + V \triangle^2 g_{00}^{(4)} +
                            \\
                            \\
                            & & & + 4 V ( 2 V - U ) \triangle^2
                            U - 24 V \triangle^2 V - 4 V \triangle U \triangle \lp U - V \rp - 12 V \lp \triangle V \rp^2 +
                            \\
                            \\
                            & & & + 2 g^{(3) \, 0i} \partial_0 \partial_i \lp 3
                            \triangle V - \triangle U \rp + \delta^{ij} \left[ - \lp \frac{1}{2} \partial_i g_{00}^{(4)} -
                            \partial_0 g_{0i}^{(3)} \rp \partial_j \lp 3 \triangle V - \triangle U \rp + \right.
                            \\
                            \\
                            & & & + \partial_i ( U - V ) \lp \frac{1}{2} \partial_j \triangle g_{00}^{(4)} -
                            2 \partial_j V \, \lp 3 \triangle V - \triangle U \rp + \partial_j \partial_0 \partial_0 U - 2
                            \partial_j U \, \triangle U - 2 U \partial_j \triangle U \rp -
                            \\
                            \\
                            & & & \left. - 2 \lp ( 2 V - U ) \partial_i U + V \partial_i V \rp \, \partial_j \lp 3
                            \triangle V - \triangle U \rp - 16 V \partial_i U \partial_j \triangle U + \partial_i
                            g_{0j}^{(3)} \partial_0 \lp 3 \triangle V - \triangle U \rp \right] -
                            \\
                            \\
                            & & & - \delta^{ij} \delta^{mn} \left[ 8 V \partial_i \partial_m U \, \partial_j
                            \partial_n U + \frac{1}{2} \lp 2 \partial_i g_{jn}^{(4)} - \partial_n g_{ij}^{(4)} \rp
                            \partial_m \lp 3 \triangle V - \triangle U \rp + \right.
                            \\
                            \\
                            & & & \left. \left. + g_{im}^{(4)} \partial_j \partial_n \lp 3
                            \triangle V - \triangle U \rp + 4 \partial_i ( U - V ) \partial_m U \partial_j \partial_n U
                            \right] \right\} +
                            \\
                            \\
                            & + & f'''(0) & \left\{ - \frac{3}{2} \lp 3 \triangle V - \triangle U \rp \triangle^2
                            g_{00}^{(4)} + 3 \triangle^2 U \left[ \frac{1}{2} \triangle g_{00}^{(4)} + \partial_0
                            \partial_0 U - 2 ( U + 3 V ) \lp 3 \triangle V - \triangle U \rp - \right. \right.
                            \\
                            \\
                            & & & \left. - 2 \delta^{ij} \partial_i U
                            \partial_j U - 2 U \triangle U \right] - \frac{37}{6} \lp \triangle U \rp^3 + \frac{51}{2} \lp \triangle U \rp^2
                            \triangle V - \frac{81}{2} \triangle U \lp \triangle V \rp^2 + \frac{117}{2} \lp \triangle V
                            \rp^3 +
                            \\
                            \\
                            & & & + 9 \triangle^2 V \left[ - \frac{1}{2} \triangle g_{00}^{(4)} - \partial_0 \partial_0 U
                            + 6 V \lp 3 \triangle V - \triangle U \rp + 2 \delta^{ij} \partial_i U \partial_j U + 2 U
                            \triangle U \right] -
                            \\
                            \\
                            & & & + 3 \lp 3 \triangle V - \triangle U \rp \partial_0 \partial_0 \lp 3 \triangle V - 2
                            \triangle U \rp + 3 \partial_0 \lp 3 \triangle V - \triangle U \rp \, \partial_0 \lp 3
                            \triangle V - \triangle U \rp +
                            \\
                            \\
                            & & & + 3 \delta^{ij} \left[ 8 \lp 3 \triangle V - \triangle U \rp \partial_i U
                            \partial_j \triangle U + \left( - \partial_i \triangle g_{00}^{(4)} + \lp 3 \triangle V -
                            \triangle U \rp \partial_i \lp U + 7 V \rp + \right. \right.
                        \end{array}
                    \end{equation*}
                    \begin{equation}\label{8fr7}
                        \begin{array}{llcl}
                            & & & \left. \left. + 6 V \partial_i \lp 3 \triangle V - \triangle U
                            \rp - 2 \partial_i \partial_0 \partial_0 U + 4 \triangle U \partial_i U
                            + 4 U \partial_i \triangle U \rp \partial_j \lp 3 \triangle V - \triangle U \right) \right] +
                            \\
                            \\
                            & & & \left. + 12 \delta^{ij} \delta^{mn} \left[ \lp 3 \triangle V - \triangle U \rp \partial_i \partial_m
                            U \, \partial_j \partial_n U + 2 \partial_i U \partial_m \lp 3 \triangle V - \triangle U \rp
                            \partial_j \partial_n U \right] \right\} -
                            \\
                            \\
                            & - & 3 f''''(0) & \lp 3 \triangle V - \triangle U \rp \left\{ \partial_i \lp 3 \triangle
                            V - \triangle U \rp \partial_j \lp 3 \triangle V - \triangle U \rp + \frac{1}{2} \lp 3
                            \triangle V - \triangle U \rp \lp 3 \triangle^2 V - \triangle^2 U \rp \right\} =
                            \\
                            \\
                            & = & 2 \kappa^2 & \left\{ \rho \left[ - \textbf{v}^2 \lp U + 2 V \rp + g_{00}^{(4)} + 2
                            g_{0i}^{(3)} v^i - 2 U \Pi \right] - 2 U p
                            \right\}.
                        \end{array}
                    \end{equation}

                Summing up, the Newtonian limit of $f(R)$ modified
                gravity theories is given by Eq.(\ref{8fr4}) and Eq.(\ref{8fr5}); the
                PN limit is given by Eq.(\ref{8fr4b}) and Eq.(\ref{8fr6}); and the
                PPN limit is given by Eq.(\ref{8fr4c}) and
                Eq.(\ref{8fr7}), respectively.

        \subsection{$f(\mathcal G)$ gravity}

            Let us consider now the following action:
                \begin{equation}\label{8fg1}
                    S=\int d^4x\sqrt{-g} \left\{ \frac{1}{2\kappa^2} \left[ R + f(\mathcal{G}) \right] +
                    \mathcal{L}_\mathrm{matter} \right\},
                \end{equation}
                where standard GR is obviously recovered for $f({\cal G})\rightarrow 0$. We proceed in the same way as for $f(R)$-gravity, but in this case we have
            $f(R, \mathcal G) \rightarrow R + f(\mathcal G)$.

            The field equations can be directly obtained from Eq. (\ref{8g2}):
                \begin{equation}\label{8fg2}
                    - \frac{1}{2} g_{\mu \nu} \lp R + f(\mathcal G) \rp + R_{\mu \nu} + 2 f'(\mathcal G) R R_{\mu \nu} -
                    4 f'(\mathcal G) R_{\mu \rho}
                    R_{\nu}^{\ \, \rho} + 2 f'(\mathcal G) R_{\alpha \beta \rho \mu} R^{\alpha \beta
                    \rho}_{\ \ \ \ \nu} - 4 f'(\mathcal G) R_{\mu \rho \nu \sigma} R^{\rho \sigma} +$$
                    $$+ 2 g_{\mu \nu} R
                    \nabla^2 f'(\mathcal G) - 4 g_{\mu \nu} R_{\rho \sigma} \nabla^\rho
                    \nabla^\sigma f'(\mathcal G) - 2 R \nabla_\mu \nabla_\nu f'(\mathcal G)
                    - 4 R_{\mu \nu} \nabla^2 f'(\mathcal G) + 4 R_{\nu \rho} \nabla^\rho \nabla_\mu f'(\mathcal G) + 4
                    R_{\mu \rho} \nabla^\rho
                    \nabla_\nu f'(\mathcal G) +$$
                    $$+ 4 R_{\mu \rho \nu \sigma} \nabla^\rho \nabla^\sigma
                    f'(\mathcal G) = 2 \kappa^2 T_{\mu \nu},
                \end{equation}
            while the trace equation is obtained from Eq.(\ref{8g3}):
                \begin{equation}\label{8fg3}
                     - R - 2 f(\mathcal G) + 2 f'(\mathcal G) \mathcal G + 2 R \nabla^2 f'(\mathcal G) - 4 R_{\rho \sigma}
                    \nabla^\rho \nabla^\sigma f'(\mathcal G) = 2 \kappa^2 T.
                \end{equation}

            From the lowest order we obtain again: $f(0) = 0$.

            \subsubsection{The Newtonian limit}

                In this case, the $(0,0)$-field equation obtained from Eq.(\ref{8fe5}) is given by:
                    \begin{equation}\label{8fg4}
                        \triangle U + 3 \triangle V = - 4 \kappa^2 \rho.
                    \end{equation}
                While the trace equation can be obtained from Eq.(\ref{8fe9}) and it reduces to:
                    \begin{equation}\label{8fg7}
                        \triangle U - 3 \triangle V = 2 \kappa^2 \rho.
                    \end{equation}

            \subsubsection{The Post Newtonian limit}

                The PN limit for the $(0,0)$-field equation can be calculated from Eq.(\ref{8fe7}) and it reduces to:
                    \begin{equation}\label{8fg5}
                        \begin{array}{lll}
                        & & \frac{1}{4} \triangle g_{00}^{(4)} + 3 ( U + V ) \triangle V + V \triangle U +
                        \frac{1}{2} \partial_0 \partial_0 U - \delta^{ij} \partial_i U \partial_j U +
                        \\
                        \\
                        & + & f'(0) \left\{ - \frac{1}{2} \lp \triangle U - \triangle V \rp^2 + 2 \delta^{im} \delta^{jn}
                        \partial_i \partial_j \lp U - V \rp \partial_m \partial_n \lp U - V \rp \right\} = 2 \kappa^2 \rho
                        \lp \Pi + \textbf{v}^2 - 4 U \rp
                        \end{array}
                    \end{equation}
                For the case of the trace equation, Eq.(\ref{8fe11}) reduces to:
                    \begin{equation}\label{8fg8}
                        \begin{array}{lll}
                        & - & \frac{1}{2} \triangle g_{00}^{(4)} + 2 V \lp \triangle U - 3 \triangle V
                        \rp + \partial_0 \partial_0 U - 2 \delta^{ij} \lp \partial_i U \rp \lp \partial_j U \rp - 2 U
                        \triangle U = 2 \kappa^2 \left\{ \rho \lp \Pi - 2 U \rp - 3 p \right\}
                        \end{array}
                    \end{equation}

            \subsubsection{The Parameterized Post Newtonian limit}

                The PPN limit for the $(0,0)$-field equation is obtained from Eq.(\ref{8fe7}) and can be written as:
                    \begin{equation*}
                        \begin{array}{lcl}
                            & & \frac{1}{2} \lp R^{(6)}_{00} + \delta^{ij} R^{(6)}_{ij} \rp - \delta^{ij} g^{(3)}_{0i}
                            R^{(3)}_{0j} - \frac{1}{2} \triangle V \left[ 3 \lp g^{(4)}_{00} + 4 U V - 4 V^2 \rp +
                            \delta^{ij} g^{(4)}_{ij} \right] +
                            \\
                            \\
                            & & + \delta^{ij} \left[ \partial_i U \partial_j ( U - V ) + 2
                            \partial_i V \partial_j ( 3 V + U ) \right] + 8 V \triangle V + 2 U \triangle U -
                            \\
                            \\
                            & & - ( U + V ) \left[ \frac{1}{2} \lp 6 \partial_0 \partial_0 V - \triangle g^{(4)}_{00} + 2
                            \delta^{ij} \lp \triangle g^{(4)}_{ij} + \partial_0 \partial_i g^{(3)}_{0j} \rp - 2 \delta^{ij}
                            \delta^{mn} \partial_i \partial_m g^{(4)}_{jn}  \rp \right] +
                            \\
                            \\
                            + & f'(0) & \left\{ - \frac{1}{2} \mathcal G^{(6)} + ( 8 V - 7 U ) \lp \triangle
                            U \rp^2 + U \lp \triangle V \rp^2 + 2 ( 4 V - 3 U ) \triangle U \triangle V + 2 \triangle U
                            \left[ \triangle g^{(4)}_{00} + 2 \partial_0 \partial_0 ( U + 2 V ) - \right. \right.
                            \\
                            \\
                            & & \left. - 4 \delta^{ij} \partial_i U \partial_j ( U + V ) \right] + \triangle V \left[
                            \triangle g^{(4)}_{00} + 6 \partial_0 \partial_0 ( U + 2 V ) + 4 \delta^{ij} \lp \partial_0
                            \partial_i g^{(3)}_{0j} + \partial_i U \partial_j ( U - 2 V ) \rp \right] +
                            \\
                            \\
                            & & + \delta^{ij} \left[ \lp \triangle g^{(3)}_{0i} + 4 \partial_0 \partial_i V \rp \triangle
                            g^{(3)}_{0j} - 8 \partial_0 \partial_i V \partial_0 \partial_j V \right] + 4 \delta^{ij}
                            \delta^{mn} \left[ U \partial_i \partial_m ( U + V ) \partial_j \partial_n ( U + V ) - \right.
                            \\
                            \\
                            & & - \partial_i \partial_m g^{(3)}_{0j} \partial_0 \partial_n V + \frac{1}{2} \left[ \partial_0
                            \lp \partial_n g^{(3)}_{0j} + \partial_j
                            g^{(3)}_{0n} - 2 \delta_{jn} \partial_0 V \rp + \delta^{kl} \lp \partial_k \lp \partial_j
                            g^{(4)}_{ln} + \partial_n g^{(4)}_{lj} \rp - \partial_j \partial_n g^{(4)}_{kl} \rp
                            - \right.
                            \\
                            \\
                            & & - \left. \triangle g^{(4)}_{jn} - \partial_j \partial_n g^{(4)}_{00} \right] - 2 V \partial_j
                            \partial_n V - ( U + 2 V ) \partial_j \partial_n U + \partial_j U \, \partial_n ( U - V )
                            - \partial_j V \, \partial_n ( 3 V + U ) -
                            \\
                            \\
                            & & \left. \left. - \delta_{jn} \lp \delta^{kl} \partial_l V \partial_k
                            ( U + V ) + 2 V \triangle V \rp \right]  \frac{1}{2} \delta^{ij} \delta^{mn} \delta^{kl} \partial_k \lp \partial_m
                            g_{0i}^{(3)} + \partial_i g_{0m}^{(3)} \rp \partial_l \lp \partial_n g_{0j}^{(3)} + \partial_j
                            g_{0n}^{(3)} \rp \right\} +
                        \end{array}
                    \end{equation*}
                    \begin{equation}\label{8fg6}
                        \begin{array}{lcl}
                            + & f''(0) & \left\{ 12 \lp \triangle U + \triangle V \rp^2
                            \triangle^2 U + 4 \lp 3 \triangle V - \triangle U \rp \lp \triangle U + \triangle V \rp
                            \triangle^2 V + \right.
                            \\
                            \\
                            & & + 4 \lp \triangle U + \triangle V \rp \delta^{ij} \left[ 3 \partial_i \triangle
                            U \partial_j \lp \triangle U + 2 \triangle V \rp - \partial_i \triangle V \partial_j \triangle V \right] +
                            \\
                            \\
                            & & + 4 \delta^{ij}
                            \delta^{mn} \left[ \triangle \left( \partial_i \partial_m ( U + V ) \partial_j \partial_n ( U
                            + V ) \right) + \partial_i \partial_m U \partial_j \partial_n \left( - 3 \lp \triangle U \rp^2
                            + \lp \triangle V \rp^2 - 6 \triangle U \triangle V + \right. \right.
                            \\
                            \\
                            & & \left. \left. \left. + 4 \delta^{kl} \delta^{rs} \partial_k \partial_r ( U + V )
                            \partial_l \partial_s ( U + V ) \right) \right] \right\} =
                            \\
                            \\
                            = & 2 \kappa^2 & \left\{ \rho
                            \left[ \textbf{v}^4 - 3 U \textbf{v}^2 + 4 U^2 + 2 g_{00}^{(4)} + 2 g_{0i}^{(3)} v^i + \Pi (
                            \textbf{v}^2 - 4 U ) \right] + p \lp \textbf{v}^2 - 2 U \rp \right\}\,.
                        \end{array}
                    \end{equation}
                And for the trace equation, Eq.(\ref{8fe13}) reduces to:
                    \begin{equation}\label{8fg9}
                        - R^{(6)} + 2 f''(0) \lp \triangle U - \triangle V \rp \triangle \left\{
                        - 3 \lp \triangle U \rp^2 + \lp \triangle V \rp^2 - 6 \lp \triangle U \rp \lp \triangle V \rp
                        + 4 \delta^{im} \delta^{jn} \partial_i \partial_j \lp U + V \rp \, \partial_m \partial_n \lp U
                        + V \rp \right\} =$$
                        $$ = 2 \kappa^2 \left\{ \rho \left[ - \textbf{v}^2 \lp U + 2 V \rp + g_{00}^{(4)} + 2 g_{0i}^{(3)}
                        v^i - 2 U \Pi \right] - 2 U p \right\}\,.
                    \end{equation}

                Summarizing, the Newtonian limit of $f(\mathcal{G})$-
                gravity theories is given by Eq.(\ref{8fg4}) and Eq.(\ref{8fg7}); the
                PN limit is given by Eq.(\ref{8fg5}) and Eq.(\ref{8fg8}); and the
                PPN limit is given by Eq.(\ref{8fg6}) and
                Eq.(\ref{8fg9}), respectively.

%%%%%%%%%%%%%%%%%%%%%
\section{Conclusions}
%%%%%%%%%%%%%%%%%%%%%
Achieving the weak field limit is the straightforward approach to compare any theory of gravity with GR. In a wide sense, this is the main consistency check in order to establish if a given theory of gravity can reproduce or not the classical experimental tests of GR and then address further phenomena and/or anomalous experimental data. 

This philosophy has recently become a paradigm due to the fact that alternative theories of gravity and extended theories of gravity  are aimed to reproduce GR results from one side (Solar System and laboratory scales) and address a huge amount of phenomena starting from quantization and renormalization of gravitational interaction (ultraviolet scales) up to the large scale structure and the cosmological accelerated behavior of the Hubble flow (infrared scales). 

The weak field limit approach is based on the theory of perturbations in terms of powers $c^{-2}$ where $c$ is the  speed of light. Newtonian, Post Newtonian and further limits strictly depends on the accuracy of such a development and on the  identification of suitable parameters that have to be confronted with the experiments.

Here we have considered a wide class of these theories, specifically the  $f(R,\mathcal G)$ gravity consisting of analytic models that are functions of the Ricci scalar
$R$ and the Gauss-Bonnet invariant ${\cal G}$. In this context, we have worked out the Newtonian, the Post Newtonian and the Parameterized Post Newtonian limits.

The main result of the approach is that new features come out with respect to GR and, due to fourth-order field equations, at least two gravitational potentials have to be considered. This fact could be extremely important in order to deal with realistic self-gravitating structures since they could results more stable and without singularities. 

Specifically, we found some general solutions in the Newtonian limit and developed in details (Newtonian, Post Newtonian e Parameterized Post Newtonian limits) the  cases of $f(R)$ gravity and $R+f({\cal G})$ gravity. In particular, the presence of the topological terms seems relevant to remove singularities and give rise to stable structures. In a forthcoming paper, we will phenomenologically confront these theoretical results with experimental data.

    \medskip
    \noindent {\bf Acknowledgements.} MDL  acknowledges the support of INFN Sez. di Napoli (Iniziative
Specifiche TEONGRAV and QGSKY).
AJLR acknowledges a JAE fellowship from CSIC (Barcelona, Spain).

\newpage
    
 \appendix
    \section{The Post Newtonian limit}\label{A1}
 In this appendix, we present  the calculations needed to obtain the PN limit for the field equations and the trace equation.  First, we write some calculations that will be needed after:

            \begin{equation*}
\begin{array}{lll}
                    \lp \nabla^2 f^* (R, \mathcal G) \rp^{(2)} & = & g^{(0) \, ij} f^*_R(0,0) \partial_i \partial_j R^{(2)}
                    \\
                    \\
                    \lp \nabla^2 f^* (R, \mathcal G) \rp^{(4)} & = & g^{(0) \, 00} \left\{ f^*_R(0,0) \partial_0
                    \partial_0 R^{(2)} - \Gamma^{(2) \, i}_{\ \ \ \ 00} f^*_R(0,0) \partial_i R^{(2)} \right\} +
                    \\
                    \\
                    & & + g^{(0) \, ij} \left\{ f^*_{RR}(0,0) \partial_i R^{(2)} \partial_j R^{(2)} + f^*_R(0,0) \partial_i
                    \partial_j R^{(4)} + f^*_{RR}(0,0) R^{(2)} \partial_i \partial_j R^{(2)} +  \right.
                    \\
                    \\
                    & & \left. + f^*_{\mathcal G}(0,0) \partial_i
                    \partial_j \mathcal G^{(4)} - \Gamma^{(2) \, k}_{\ \ \ \ ij} f^*_R(0,0) \partial_k R^{(2)} \right\} +
                    g^{(2) \, ij} f^*_R(0,0) \partial_i \partial_j R^{(2)}
                    \\
                    \\
\end{array}
            \end{equation*}

            \begin{equation*}
                \begin{array}{lll}
                    & & - 2 \delta^{ij} \delta^{mn} \partial_i
                    \partial_m g_{0j}^{(3)} \partial_0 \partial_n V
                    + 2 \delta^{ij} \partial_0 \partial_i V \lp \triangle g_{0j}^{(3)} - 2 \partial_0 \partial_j V \rp
                    \\
                    \\
                    \lp R_{0 \rho 0 \sigma} R^{\rho \sigma} \rp^{(4)} & = & - \lp \triangle U \rp \lp \triangle V \rp
                    \\
                    \\
                    \lp R_{0 \rho 0 \sigma} R^{\rho \sigma} \rp^{(6)} & = & \triangle V
                    \left\{ \frac{1}{2} \triangle
                    g^{(4)}_{00} - 3 \, \partial_0 \partial_0 V + 4 V \triangle U - \delta^{ij} \left[ \frac{1}{2}
                    \partial_0 \lp \partial_i g^{(3)}_{0j} + \partial_j g^{(3)}_{0i} \rp + 2 \partial_i U \, \partial_j
                    ( 2 U - V ) \right] \right\} +
                    \\
                    \\
                    & & + \delta^{ij} \delta^{mn} \partial_i \partial_m U \, \left\{ \frac{1}{2} \left[ \partial_0 \lp
                    \partial_n g^{(3)}_{0j} + \partial_j g^{(3)}_{0n} + 2 \delta_{jn} \partial_0 V \rp - \right. \right.
                    \\
                    \\
\end{array}
            \end{equation*}

        For the $(0,0)$-field equation the following results are necessary:
            \begin{equation*}
                \begin{array}{cll}
                    \lp - \frac{1}{2} g_{00} f(R, \mathcal G) \rp^{(4)} & = & - \frac{1}{2} g_{00}^{(4)} f(0,0) -
                    \frac{1}{2} g_{00}^{(2)} f_R(0,0) R^{(2)} -
                    \\
                    \\
                    & & - \frac{1}{2} g_{00}^{(0)} \lp \frac{1}{2} f_{RR}(0,0)
                    R^{(2) \, 2} + f_R(0,0) R^{(4)} + f_{\mathcal G}(0,0) \mathcal G^{(4)} \rp
                    \\
                    \\
                    \lp f_R (R, \mathcal G) R_{00} \rp^{(4)} & = & f_R(0,0) R_{00}^{(4)} + f_{RR}(0,0) R^{(2)} R_{00}^{(2)}
                    \\
                    \\
                    \lp g_{00} \nabla^2 f_R (R, \mathcal G) \rp^{(4)} & = & g_{00}^{(2)} g^{(0) \, ij} f_{RR}(0,0)
                    \partial_i \partial_j R^{(2)} + g_{00}^{(0)} \left[ g^{(0) \, 00} f_{RR}(0,0) \lp \partial_0
                    \partial_0 R^{(2)} - \Gamma^{(2) \, i}_{\ \ \ \, 00} \partial_i R^{(2)} \rp
                    + \right.
                    \\
                    \\
\end{array}
            \end{equation*}
            \begin{equation*}
                \begin{array}{cll}
                    \lp - 4 f_{\mathcal G}(R, \mathcal G) R_{0 \rho 0 \sigma} R^{\rho
                    \sigma} \rp^{(4)} & = & - 4
                    f_{\mathcal G}(0,0) \lp R_{0 \rho 0 \sigma} R^{\rho \sigma} \rp^{(4)}
                    \\
                    \\
                    \lp 2 g_{00} R \nabla^2 f_{\mathcal G}(R, \mathcal G) \rp^{(4)} & = & 2 g_{00}^{(0)} R^{(2)} g^{(0) \, ij}
                    f_{\mathcal G R}(0,0) \partial_i \partial_j R^{(2)}
                    \\
                    \\
                    \lp - 4 g_{00} R_{\rho \sigma} \nabla^\rho \nabla^\sigma f_{\mathcal G}(R, \mathcal G) \rp^{(4)} & = & - 4
                    g_{00}^{(0)} R_{kl}^{(2)} g^{(0) \, ik} g^{(0) \, jl} f_{\mathcal G R}(0,0) \partial_i \partial_j R^{(2)}
                    \\
                    \\
                    \lp - 2 R \nabla_0 \nabla_0 f_{\mathcal G}(R, \mathcal G) \rp^{(4)} & = & 0
                    \\
                    \\
                    \lp - 4 R_{00} \nabla^2 f_{\mathcal G}(R, \mathcal G) \rp^{(4)} & = & - 4 R_{00}^{(2)} g^{(0) \, ij}
                    f_{\mathcal G R}(0,0) \partial_i \partial_j R^{(2)}
                    \\
                    \\
                    \lp 8 R_{0 \rho} \nabla^\rho \nabla_0 f_{\mathcal G}(R, \mathcal G) \rp^{(4)} & = & 0
                    \\
                    \\
                    \lp 4 R_{0 \rho 0 \sigma} \nabla^\rho \nabla^\sigma f_{\mathcal G}(R, \mathcal G) \rp^{(4)} & = &
                    4 R_{0 k 0 l}^{(2)} g^{(0) \, ik} g^{(0) \, jl} f_{\mathcal G R}(0,0) \partial_i \partial_j R^{(2)}
                    \\
                    \\
                    \lp 2 \kappa^2 T_{00} \rp^{(4)} & = & 2 \kappa^2 T_{00}^{(2)}
                \end{array}
            \end{equation*}

        In the case of the trace equation, we need:
            \begin{equation*}
                \begin{array}{cll}
                    \lp - 2 f(R, \mathcal G) \rp^{(4)} & = & - 2 f_R(0,0) R^{(4)} - f_{RR}(0,0) R^{(2) \, 2} - 2
                    f_{\mathcal G}(0,0) \mathcal G^{(4)}
                    \\
                    \\
                    \lp f_R (R, \mathcal G) R \rp^{(4)} & = & f_R(0,0) R^{(4)} + f_{RR}(0,0) R^{(2) \, 2}
                    \\
                    \\
                    \lp 3 \nabla^2 f_R (R, \mathcal G) \rp^{(4)} & = & 3 f_{RR}(0,0) \left\{ g^{(0) \, 00} \lp \partial_0
                    \partial_0 R^{(2)} - \Gamma^{(2) \, i}_{\ \ \ \ 00} \partial_i R^{(2)} \rp + g^{(0) \, ij} \lp
                    \partial_i \partial_j R^{(4)} - \Gamma^{(2) \, k}_{\ \ \ \, ij}
                    \partial_k R^{(2)} \rp + \right.
                    \\
                    \\
                    & & \left. + g^{(2) \, ij} \partial_i \partial_j R^{(2)} \right\} + 3 f_{RRR}(0,0) g^{(0) \, ij}
                    \left\{ \partial_i R^{(2)} \partial_j R^{(2)} + R^{(2)} \partial_i \partial_j R^{(2)} \right\} +
                    \\
                    \\
\end{array}
            \end{equation*}

    \section{The Parameterized Post Newtonian Limit}\label{A2}
We present now the calculations needed to obtain the PPN limit for the field equations and the trace equation.
 In the case of the $(0,0)$-field equation we need to know the following expressions:
            \begin{equation*}
                \begin{array}{cll}
                    \lp - \frac{1}{2} g_{00} f(R, \mathcal G) \rp^{(6)} & = & - \frac{1}{2} g_{00}^{(6)} f(0,0) -
                    \frac{1}{2} f_R(0,0) \left\{ g_{00}^{(4)} R^{(2)} + g_{00}^{(2)} R^{(4)} + g_{00}^{(0)} R^{(6)}
                    \right\} -
                    \\
                    \\
                    & & - \frac{1}{2} f_{\mathcal G}(0,0) \left\{ g_{00}^{(2)} \mathcal G^{(4)} + g_{00}^{(0)}
                    \mathcal G^{(6)} \right\} - \frac{1}{2} f_{RR}(0,0) \left\{ \frac{1}{2} g_{00}^{(2)} R^{(2) \, 2} + g_{00}^{(0)}
                    R^{(2)} R^{(4)} \right\}
                    \\
                    \\
                    & & - \frac{1}{2} g_{00}^{(0)} f_{R \mathcal G}(0,0) R^{(2)}
                    \mathcal G^{(4)} - \frac{1}{12} f_{RRR}(0,0) R^{(2) \, 3}
                    \\
                    \\
                    \lp f_R (R, \mathcal G) R_{00} \rp^{(6)} & = & f_R(0,0) R_{00}^{(6)} + f_{RR}(0,0) \left\{ R^{(2)}
                    R_{00}^{(4)} + R^{(4)} R_{00}^{(2)} \right\} + f_{R \mathcal G}(0,0) \mathcal G^{(4)} R_{00}^{(2)} +
                    \\
                    \\
\end{array}
            \end{equation*}

            \begin{equation*}
                \begin{array}{cll}
                    \lp 2 f_{\mathcal G}(R, \mathcal G) R R_{00} \rp^{(6)} & = & 2 f_{\mathcal G}(0,0) \left\{ R^{(2)}
                    R^{(4)}_{00} + R^{(4)} R^{(2)}_{00} \right\} + 2 f_{R \mathcal G}(0,0) R^{(2) \, 2}
                    R^{(2)}_{00}
                    \\
                    \\
                    \lp - 4 f_{\mathcal G}(R, \mathcal G) g^{\rho \sigma} R_{0 \rho} R_{0 \sigma} \rp^{(6)} & = & - 4
                    f_{\mathcal G}(0,0) \left\{ 2 g^{(0) \, 00} R^{(2)}_{00} R^{(4)}_{00} + g^{(2) \, 00} R^{(2) \, 2}_{00}
                    + g^{(0) \, ij} R^{(3)}_{0i} R^{(3)}_{0j} \right\} -
                    \\
                    \\
                    & & - 4 f_{R \mathcal G}(0,0) g^{(0) \, 00} R^{(2)}
                    R^{(2) \, 2}_{00}
                    \\
                    \\
                    \lp 2 f_{\mathcal G}(R, \mathcal G) R_{\alpha \beta \rho 0} R^{\alpha \beta \rho}_{\ \ \ \ 0} \rp^{(6)}
                    & = & 2 f_{\mathcal G}(0,0) \lp R_{\alpha \beta \rho 0} R^{\alpha \beta \rho}_{\ \ \ \ 0}
                    \rp^{(6)} + 2 f_{R \mathcal G}(0,0) R^{(2)} \lp R_{\alpha \beta \rho 0}
                    R^{\alpha \beta \rho}_{\ \ \ \ 0} \rp^{(4)}
                    \\
                    \\
\end{array}
            \end{equation*}

        Finally, in the case of the trace equation, the expressions needed are:
            \begin{equation*}
                \begin{array}{cll}
                    \lp - 2 f(R, \mathcal G) \rp^{(6)} & = & - 2 \left\{ f_R(0,0) R^{(6)} + f_{\mathcal G}(0,0)
                    \mathcal G^{(6)} + f_{RR}(0,0) R^{(2)} R^{(4)} + f_{R \mathcal G}(0,0) R^{(2)} \mathcal G^{(4)}
                    + \right.
                    \\
                    \\
                    & & + \left. \frac{1}{6} f_{RRR}(0,0) R^{(2) \, 3} \right\}
                    \\
                    \\
                    \lp f_R (R, \mathcal G) R \rp^{(6)} & = & f_R(0,0) R^{(6)} + 2 f_{RR}(0,0) R^{(2)} R^{(4)} +
                    f_{R \mathcal G}(0,0) R^{(2)} \mathcal G^{(4)} + \frac{1}{2} f_{RRR}(0,0) R^{(2) \, 3}
                    \\
                    \\
                    \lp 3 \nabla^2 f_R (R, \mathcal G) \rp^{(6)} & = & 3 \lp \nabla^2 f_R (R, \mathcal G) \rp^{(6)}
                    \\
                    \\
                    \lp 2 f_{\mathcal G}(R, \mathcal G) \mathcal G \rp^{(6)} & = & 2 f_{\mathcal G}(0,0) \mathcal G^{(6)}
                    + 2 f_{R \mathcal G}(0,0) R^{(2)} \mathcal G^{(4)}
                    \\
                    \\
                    \lp 2 R \nabla^2 f_{\mathcal G}(R, \mathcal G) \rp^{(6)} & = & 2 R^{(2)} \lp \nabla^2
                    f_{\mathcal G}(R, \mathcal G) \rp^{(4)} + 2 R^{(4)} \lp \nabla^2 f_{\mathcal G}(R, \mathcal G)
                    \rp^{(2)}
                \end{array}
            \end{equation*}
            \begin{equation*}
                \begin{array}{cll}
                    \lp - 4 R_{\rho \sigma} \nabla^\rho \nabla^\sigma f_{\mathcal G}(R, \mathcal G) \rp^{(6)} & = & - 4
                    R^{(2)}_{\rho \sigma} \lp \nabla^\rho \nabla^\sigma f_{\mathcal G}(R, \mathcal G) \rp^{(4)} - 4
                    R^{(3)}_{\rho \sigma} \lp \nabla^\rho \nabla^\sigma f_{\mathcal G}(R, \mathcal G) \rp^{(3)} - 4
                    R^{(4)}_{\rho \sigma} \lp \nabla^\rho \nabla^\sigma f_{\mathcal G}(R, \mathcal G) \rp^{(2)}
                    \\
                    \\
                    \lp 2 \kappa^2 T \rp^{(6)} = 2 \kappa^2 T^{(4)} & = & 2 \kappa^2 \lp g^{(0) \, 00} T^{(4)}_{00} + g^{(2) \, 00}
                    T^{(2)}_{00} + g^{(4) \, 00} T^{(0)}_{00} + g^{(3) \, 0i} T^{(1)}_{0i} + g^{(0) \, ij} T^{(4)}_{ij} +
                    g^{(2) \, ij} T^{(2)}_{ij} \rp
                \end{array}
            \end{equation*}

 %%%%%%%%%%%%%%%%%%


\begin{thebibliography}{10}
%%%%%%%%%%%%%%%%%%

\bibitem{PhysRepnostro}S. Capozziello, M. De Laurentis, {\it Phys. Rept.} {\bf 509}, 167 (2011).

\bibitem{OdintsovPR} S. Nojiri, S. D. Odintsov, {\it Phys. Rept.} {\bf 505}, 59 (2011).

\bibitem{Mauro}S. Capozziello, M. Francaviglia, {\it Gen. Rel. Grav.} {\bf 40}, 357, (2008).

\bibitem{Nojiri:2006ri}S. Nojiri and S. D. Odintsov, eConf C {\bf 0602061} (2006) 06 [Int.\ J.\
Geom.\ Meth.\ Mod.\ Phys.\  {\bf 4} (2007) 115].

\bibitem{DeFelice}A. De Felice,  Tsujikawa, {\it Living Rev.Rel.} {\bf 13}, 3 (2010).

\bibitem{faraoni} Capozziello S., De Laurentis M., Faraoni V., 2009 ,{\it The Open Astr. Jour} , {\bf 2}, 1874.

\bibitem{buch} H. Buchdahl, J. Phys. A \textbf{12}, 1229 (1979).

\bibitem{ker} R. Kerner, Gen. Rel. Grav. \textbf{14}, 453 (1982).

\bibitem{bar} J. D. Barrow and A. C. Ottewill, J. Phys. A \textbf{16},
  2757 (1983).

\bibitem{mag} G. Magnano, M. Ferraris and M. Francaviglia,
  Gen. Rel. Grav. \textbf{19}, 465 (1987).

  \bibitem{Stel} K. S. Stelle, Phys. Rev. D \textbf{16}, 953 (1977).

\bibitem{Star80}
Starobinsky, A.A., {\em Phys. Lett. B}, {\bf 91}, 99--102, (1980).

\bibitem{annalen}S. Capozziello, M. De Laurentis,
Annalen der Physik 524, 1 (2012);\\
I De Martino, M. De Laurentis, F. Atrio-Barandela, S. Capozziello,      arXiv:1310.0693 [astro-ph.CO]

\bibitem{ghost1}
Stelle, K.S.,  {\em Gen. Relativ.
  Gravit.}, {\bf 9}, 353--371, (1978).

\bibitem{ghost1d}
Barth, N.H., and Christensen, S.M., ``Quantizing Fourth Order Gravity Theories.
  1. The Functional Integral'', {\em Phys. Rev. D}, {\bf 28}, 1876--1893,
  (1983). {\small[\href{http://dx.doi.org/10.1103/PhysRevD.28.1876}{DOI}]}.

 \bibitem{ghost1dd}
Hindawi, A., Ovrut, B.A., and Waldram, D., ``Consistent Spin-Two Coupling and
  Quadratic Gravitation'', {\em Phys. Rev. D}, {\bf 53}, 5583--5596, (1996).
  {\small[\href{http://dx.doi.org/10.1103/PhysRevD.53.5583}{DOI}]}.

 \bibitem{ghost1ddd}
Boulanger, N., Damour, T., Gualtieri, L., and Henneaux, M., ``Inconsistency of
  interacting, multi-graviton theories'', {\em Nucl. Phys. B}, {\bf 597},
  127--171, (2001).

  \bibitem{ghost2}
N\'u\~nez, A., and Solganik, S., ``Ghost constraints on modified gravity'',
  {\em Phys. Lett. B}, {\bf 608}, 189--193, (2005).
  {\small[\href{http://dx.doi.org/10.1016/j.physletb.2005.01.015}{DOI}]}.


\bibitem{ghost2d}
Chiba, T., ``Generalized gravity and ghost'', {\em J. Cosmol. Astropart.
  Phys.}, {\bf 2005}(03), 008, (2005).
  {\small[\href{http://dx.doi.org/10.1088/1475-7516/2005/03/008}{DOI}]}.

\bibitem{ghost2dd}
Navarro, I., and Van~Acoleyen, K., ``On the Newtonian limit of Generalized
  Modified Gravity Models'', {\em Phys. Lett. B}, {\bf 622}, 1--5, (2005).
  {\small[\href{http://dx.doi.org/10.1016/j.physletb.2005.07.008}{DOI}]}.

\bibitem{Cal05}
Calcagni, G., Tsujikawa, S., and Sami, M., ``Dark energy and cosmological
  solutions in second-order string gravity'', {\em Class. Quantum Grav.}, {\bf
  22}, 3977--4006, (2005)

  \bibitem{DeFelice06}
De~Felice, A., Hindmarsh, M., and Trodden, M., ``Ghosts, instabilities, and
  superluminal propagation in modified gravity models'', {\em J. Cosmol.
  Astropart. Phys.}, {\bf 2006}(08), 005, (2006).
  {\small[\href{http://dx.doi.org/10.1088/1475-7516/2006/08/005}{DOI}]}.

  \bibitem{DeFelice06d}
Calcagni, G., de~Carlos, B., and De~Felice, A., ``Ghost conditions for
  Gauss-Bonnet cosmologies'', {\em Nucl. Phys. B}, {\bf 752}, 404--438, (2006).
  {\small[\href{http://dx.doi.org/10.1016/j.nuclphysb.2006.06.020}{DOI}]}.

    \bibitem{fGO}
Nojiri, S., and Odintsov, S.D., ``Modified Gauss-Bonnet theory as gravitational
  alternative for dark energy'', {\em Phys. Lett. B}, {\bf 631}, 1--6, (2005).

 \bibitem{DeTsuji1}
De~Felice, A., and Tsujikawa, S., ``Construction of cosmologically viable
  $f({\mathcal G})$ gravity models'', {\em Phys. Lett. B}, {\bf 675}, 1--8,
  (2009).

 \bibitem{DeTsuji2}
De~Felice, A., and Tsujikawa, S., ``Solar system constraints on $f({\mathcal
  G})$ gravity models'', {\em Phys. Rev. D}, {\bf 80}, 063516, (2009).

\bibitem{Cognola}
Cognola, G., Elizalde, E., Nojiri, S., Odintsov, S., and Zerbini, S.,
  ``String-inspired Gauss-Bonnet gravity reconstructed from the universe
  expansion history and yielding the transition from matter dominance to dark
  energy'', {\em Phys. Rev. D}, {\bf 75}, 086002, (2007).

  \bibitem{DeHind}
De~Felice, A., and Hindmarsh, M., ``Unsuccessful cosmology with modified
  gravity models'', {\em J. Cosmol. Astropart. Phys.}, {\bf 2007}(06), 028,
  (2007).

 \bibitem{LiMota}
Li, B., Barrow, J.D., and Mota, D.F., ``The Cosmology of Modified Gauss-Bonnet
  Gravity'', {\em Phys. Rev. D}, {\bf 76}, 044027, (2007).

\bibitem{Zhou}
Zhou, S.-Y., Copeland, E.J., and Saffin, P.M., ``Cosmological Constraints on
  $f(G)$ Dark Energy Models'', {\em J. Cosmol. Astropart. Phys.}, {\bf
  2009}(07), 009, (2009).

 \bibitem{Uddin}
Uddin, K., Lidsey, J.E., and Tavakol, R., ``Cosmological scaling solutions in
  generalised Gauss-Bonnet gravity theories'', {\em Gen. Relativ. Gravit.},
  {\bf 41}, 2725--2736, (2009).

\bibitem{Cognola:2006eg}
  G.~Cognola, E.~Elizalde, S.~'i.~Nojiri, S.~D.~Odintsov and S.~Zerbini,
  Phys.\ Rev.\ D {\bf 73} (2006) 084007.

\bibitem{rob62} H. P. Robertson, in \textit{Space Age Astronomy},
  ed. A. J. Deutsch and W. B. Klemperer, Academic, New York (1962).

\bibitem{Edd} A. S. Eddington, \textit{The Mathematical Theory of
  Relativity, 2nd ed.}, Cambridge UP, Cambridge (1924).

\bibitem{sch67} L. I. Schiff, in \textit{Relativity Theory and Astrophysics
  I. Relativity and Cosmology}, ed. J. Ehlers, American Mathematical
  Society, Providence (1967)

  \bibitem{Nor} K. Nordvedt, Phys. Rev. \textbf{169}, 1017 (1968).
  C. M. Will, Astrophys. J. \textbf{163}, 611 (1971).  C. M. Will and
  K. Nordvedt, Astrophys. J. \textbf{177}, 757 (1972).

\bibitem{tegp} C. M. Will, \textit{Theory and Experiment in
  Gravitational Physics, revised ed.}, Cambridge UP, Cambridge (1993).

\bibitem{cap}  S. Capozziello and A. Troisi, Phys. Rev. D \textbf{72},
  044022 (2005).

\bibitem{olmo}  G. J. Olmo, Phys. Rev. Lett. \textbf{95}, 261102
  (2005).

\bibitem{frst}  P. Teyssandier and P. Tourranc,
  J. Math. Phys. \textbf{24}, 2793 (1983).

\bibitem{sot}  T. P. Sotiriou and E. Barausse, Phys. Rev. D
  \textbf{75}, 084007 (2007).

\bibitem{will} Will C.M. "{\it Theory and experiment in gravitational physics}" Cambridge University Press, UK (1993).

 \bibitem{Capozziello:2010gu}S. Capozziello, A. Stabile,
          gr\,-\,qc/1009.3441 (2010).
%
         \bibitem{Stabile:2010zk}A. Stabile, Phys. Rev. D \textbf{82},
            064021 (2010).
%
 \bibitem{Capozziello:2007ms}S. Capozziello, A. Stabile,
 A. Troisi, Phys. Rev. D \textbf{76},
  104019 (2007).
%
 \bibitem{Capozziello:2009vr} S. Capozziello, A. Stabile, A. Troisi, Mod.~Phys.~Lett.~A \textbf{24}, 659 (2009).
%


\end{thebibliography}
\end{document}